\documentclass[aps,prd,reprint,superscriptaddress,nofootinbib,floatfix,longbibliography]{revtex4-1}
\usepackage[dvipsnames]{xcolor}
\usepackage{amssymb,amsmath,amsfonts,bm,slashed,soul,ragged2e,graphicx,epstopdf,hyperref,array,slashed}
\usepackage[utf8]{inputenc}
\usepackage{colortbl,color,caption,subcaption}
\enlargethispage{\baselineskip}
\definecolor{steelblue}{RGB}{25,25,112}
\definecolor{dullblue}{rgb}{0,0.298,0.49}
\definecolor{darkred}{rgb}{0.545,0,0}
\definecolor{darkorange}{RGB}{222,132,69}
\definecolor{darkgreen}{RGB}{126,171,85}
\definecolor{blue2}{cmyk}{1, 0.1, 0.1, 0}
\hypersetup{colorlinks,linkcolor={darkred},citecolor={blue},urlcolor={dullblue}}
\DeclareCaptionJustification{justified}{\justifying}
\everymath{\displaystyle} 

\usepackage{comment}
\usepackage{braket}
\usepackage{amsmath}
\usepackage{soul}
\usepackage{makecell}

\newcommand{\nn}{{\nonumber}}

\newcommand{\beq}{\begin{equation}}
\newcommand{\eeq}{\end{equation}}
\newcommand{\bea}{\begin{eqnarray}}
\newcommand{\eea}{\end{eqnarray}}

\newcommand{\gsim}{\lower.7ex\hbox{$\;\stackrel{\textstyle>}{\sim}\;$}}
\newcommand{\lsim}{\lower.7ex\hbox{$\;\stackrel{\textstyle<}{\sim}\;$}}

\newcommand{\be}{\begin{equation}}
\newcommand{\ee}{\end{equation}}
\newcommand{\ba}{\begin{eqnarray}}
\newcommand{\ea}{\end{eqnarray}}

\newcommand{\D}{\mathrm{d}}

\newcommand{\occ}{\textrm{occ}}
\newcommand{\mi}{\mathrm{i}}

\RequirePackage[normalem]{ulem}

\begin{document}

\title{Simultaneous Resonant and Broadband Detection of Ultralight Dark Matter and High-Frequency Gravitational Waves via Cavities and Circuits}

\author{Yifan Chen}
\email{Corresponding author: yifan.chen@nbi.ku.dk}
\affiliation{
Center of Gravity, Niels Bohr Institute, Blegdamsvej 17, 2100 Copenhagen, Denmark}
\author{Chunlong Li}
\affiliation{
CAS Key Laboratory of Theoretical Physics, Institute of Theoretical
Physics, Chinese Academy of Sciences, Beijing 100190, China}
\author{Yuxin Liu}
\affiliation{International Centre for Theoretical Physics Asia-Pacific, Beijing/Hangzhou, China}
\affiliation{
School of Physical Sciences, University of Chinese Academy of Sciences, Beijing 100049, China}
\author{Jing Shu}
\email{Corresponding author: jshu@pku.edu.cn}
\affiliation{School of Physics and State Key Laboratory of Nuclear Physics and Technology, Peking University, Beijing 100871, China}
\affiliation{Center for High Energy Physics, Peking University, Beijing 100871, China}
\affiliation{Beijing Laser Acceleration Innovation Center, Huairou, Beijing, 101400, China}
\author{Yuting Yang}
\affiliation{
CAS Key Laboratory of Theoretical Physics, Institute of Theoretical
Physics, Chinese Academy of Sciences, Beijing 100190, China}
\affiliation{
School of Physical Sciences, University of Chinese Academy of Sciences, Beijing 100049, China}
\author{Yanjie Zeng}
\affiliation{
CAS Key Laboratory of Theoretical Physics, Institute of Theoretical
Physics, Chinese Academy of Sciences, Beijing 100190, China}
\affiliation{
School of Physical Sciences, University of Chinese Academy of Sciences, Beijing 100049, China}

\begin{abstract}

Electromagnetic resonant systems, such as cavities and LC circuits, are widely used to detect ultralight boson dark matter and high-frequency gravitational waves. However, the narrow bandwidth of single-mode resonators necessitates multiple scan steps to cover broad frequency ranges. By incorporating a network of auxiliary modes via beam-splitter-type and non-degenerate parametric couplings, we enable broadband detection with an effective bandwidth of each scan matching the order of the resonant frequency, while maintaining a strong signal response. In heterodyne upconversion detection, where a background cavity mode transitions into another due to a potential background source, multiple orders of the source frequency can be probed with high sensitivity without tuning the cavity frequency. Consequently, our method allows for significantly deeper exploration of the parameter space within the same integration time compared to single-mode detection.

\end{abstract}
\date{\today}

\maketitle

\tableofcontents

\section{Introduction}
Axions~\cite{Preskill:1982cy, Abbott:1982af, Dine:1982ah} and dark photons~\cite{Nelson:2011sf} are compelling candidates for dark matter (DM) due to their natural prediction in the compactification of higher dimensional fundamental theories~\cite{Svrcek:2006yi, Abel:2008ai, Arvanitaki:2009fg, Goodsell:2009xc}. The QCD axion, in particular, provides a solution to the strong CP problem~\cite{Peccei:1977hh}. Experimental efforts are underway to detect these bosonic DM candidates through their electromagnetic coupling, using techniques like resonant microwave cavities~\cite{Sikivie:1983ip, Sikivie:1985yu} and superconducting circuits~\cite{Sikivie:2013laa, Chaudhuri:2014dla, Kahn:2016aff}. In both cases, axion fields in the presence of a strong background magnetic field, or dark photon fields themselves, can induce effective currents that serve as signals in haloscope experiments~\cite{Nguyen:2019xuh,Dixit:2020ymh,Ghosh:2021ard,Caputo:2021eaa,Cervantes:2022epl,Cervantes:2022gtv,SHANHE:2023kxz}.

On the other hand, gravitational waves (GW) with frequencies higher than kHz can offer insights into early universe cosmology and particle physics beyond the standard model~\cite{Aggarwal:2020olq,Aggarwal:2025noe}. Two noteworthy phenomena—the inverse Gertsenshtein effect~\cite{Gertsenshtein}, akin to axion electrodynamics, and mechanical resonance~\cite{Pegoraro:1977uv,Pegoraro:1978gv,Reece:1984gv,Berlin:2023grv}—have the capability to convert GW into photons in the presence of a background electromagnetic field. The largely unexplored parameter space of high-frequency GW (HFGW) has gained significant attention, leading to the implementation of axion haloscope experiments aimed at detecting its signatures~\cite{Herman:2020wao, Berlin:2021txa, Domcke:2022rgu, Sokolov:2022dej, Berlin:2022hfx, Tobar:2022pie, Berlin:2023grv,Navarro:2023eii}.

The scan rate, a critical figure of merit for covering broad frequency ranges~\cite{Zheng:2016qjv,Malnou:2018dxn,HAYSTAC:2020kwv,Lehnert:2021gbj}, is influenced by the trade-off between the bandwidth of each scan and its signal response during resonant detection. In single-mode resonators, responses to the signal and to the noise due to intrinsic fluctuations are identical, resulting in the effective bandwidth of each scan as the range where this intrinsic noise predominates over readout noise with a relatively flat power spectrum~\cite{Chaudhuri:2018rqn,Chaudhuri:2019ntz}. Consequently, by adjusting the readout coupling, it is possible to optimize sensitivity and approach the standard quantum limit~\cite{Krauss:1985ub,Zheng:2016qjv,Malnou:2018dxn,Chaudhuri:2018rqn,Chaudhuri:2019ntz,Berlin:2019ahk,Lasenby:2019prg,Lasenby:2019hfz,Lehnert:2021gbj}.

Exceeding the standard quantum limit in single-mode resonators involves broadening the effective bandwidth of each scan. A practical approach to achieve this is by reducing readout noise through squeezing techniques~\cite{Zheng:2016qjv, Malnou:2018dxn, HAYSTAC:2020kwv, Lehnert:2021gbj}. Recent studies have shown that a multi-mode resonant system with auxiliary non-degenerate parametric interactions can significantly increase the signal response at off-resonant frequencies, thereby considerably expanding the effective bandwidth~\cite{WLC, Chen:2021bgy, Wurtz:2021cnm}.

In this study, we demonstrate the ultimate sensitivity limit achievable by a multi-mode resonator, which allows for a substantially increased quantum limit for the scan rate. This newly derived limit enables experiments to achieve a bandwidth as extensive as the resonant frequency for each scan. Notably, by applying multi-mode generalization to heterodyne upconversion detections—where bosonic fields and HFGWs induce transitions between two quasi-degenerate modes~\cite{Berlin:2019ahk, Lasenby:2019prg, Berlin:2020vrk, Tobar:2022pie}—it becomes feasible to realize a simultaneous resonant and broadband detector. This configuration achieves a significant signal response while obtaining an effective bandwidth that spans several orders in the frequency domain.

The structure of this paper is organized as follows: Section~\ref{sec:detectors} briefly introduces resonant cavity and LC-circuit detectors, focusing on their mode quantization and response to effective currents. Section~\ref{sec:signals} discusses the effective currents induced by axions, dark photons, and HFGWs, as well as the mechanical vibration signals induced by HFGWs. Section~\ref{sec:single-mode} applies the input-output formalism to derive the response and quantum limit of single-mode resonators. Section~\ref{sec:multi-mode} explores multi-mode generalizations, presenting example configurations and detailing their responses and the enhanced response widths. Section~\ref{sec:multi-mode realization} discusses the practical realization and potential experimental challenges of these configurations. In Sec.~\ref{sec:resonant-broadband}, we apply the achieved bandwidth to three types of detection. Finally, section~\ref{sec:conclusion} summarizes the results and offers future outlooks.

\section{Resonant Electromagnetic Systems}\label{sec:detectors}
\begin{figure}[htb]
    \centering
    \includegraphics[width=0.5\textwidth]{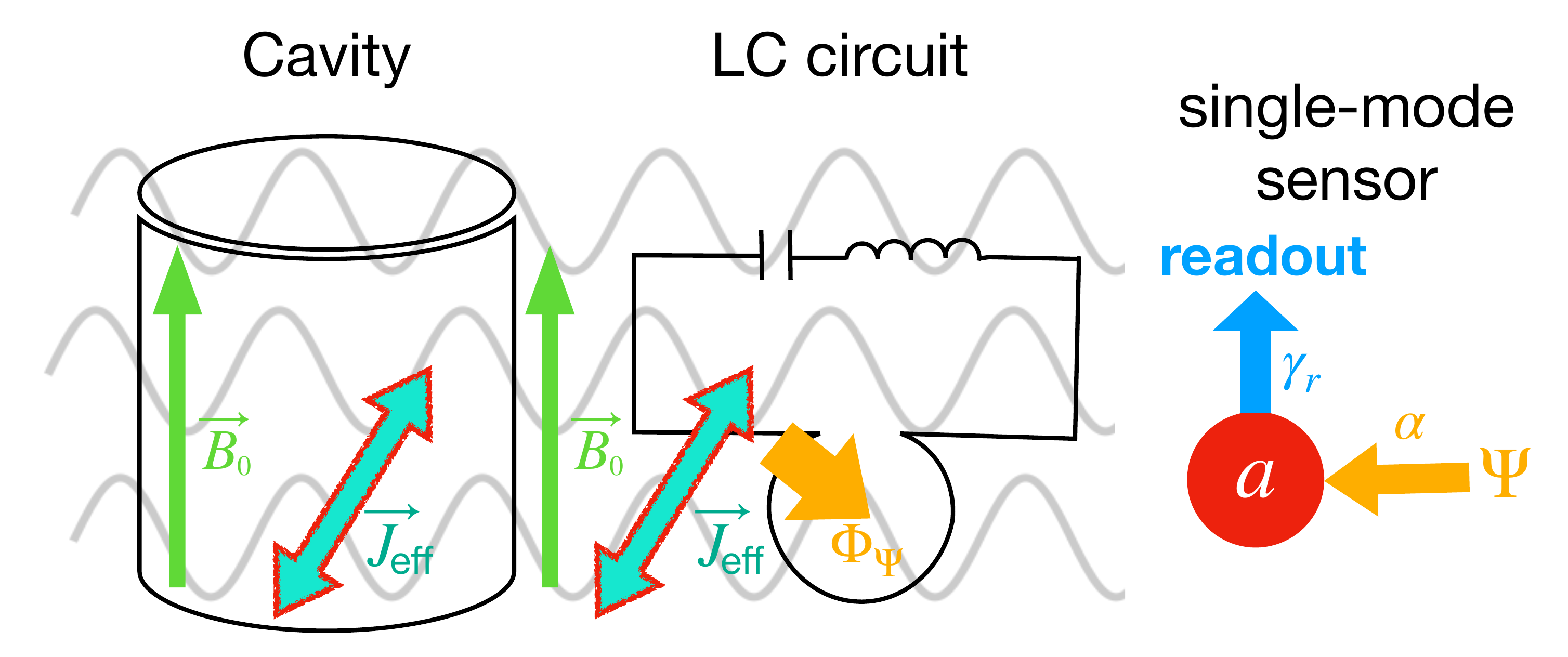}
    \caption{\textbf{Left:} An illustration depicting a resonant cavity and a superconducting LC circuit used to detect effective currents $\vec{J}_{\rm eff}$ originating from dark matter (DM) or high-frequency gravitational waves (HFGW). Background magnetic fields $\vec{B}_0$ are utilized for axion or GW detection. \textbf{Right:} A depiction of both detection scenarios unified within a single-mode resonant sensor, which probes the DM or HFGW, labeled as $\Psi$. The signal is then transmitted to the readout port.}
    \label{fig:cc}
\end{figure}

In this section, we introduce two electromagnetic resonant systems commonly used as detectors: resonant cavities and LC circuits. The left panel of Fig.~\ref{fig:cc} illustrates these detection schemes. We detail the parameterization of the interaction between quantized resonant modes and an effective current signal. A cavity mode resonates when the current’s frequency falls within the resonant bandwidth and its wave function spatially overlaps with the cavity mode. Superconducting LC circuits utilize pick-up loops to convert the magnetic flux, induced by the effective current in a shielded environment, into a voltage signal. Additionally, section~\ref{sec:circuit-representation} presents equivalent circuit representations of a cavity mode.

\subsection{Resonant Cavity}

Electromagnetic fields can be generated in the presence of an effective current, denoted as $J^\mu_{\rm eff}$. A resonant cavity, characterized by a low dissipation quality factor $Q_{\rm int}$, provides a straightforward means to amplify this weak signal. Within the cavity volume $V$, the electromagnetic fields become quantized bound states. Their wavefunctions, parameterized in the Weyl gauge, are expressed as:
follows:
\begin{equation}
    \vec{A}  = \sum_n \frac{1}{\sqrt{2\omega_{\rm rf}^n}} \hat{a}_n^\dagger \vec{\epsilon}_n(\vec{r}\,)e^{-\mi\omega_{\rm rf}^n\, t}+ h.c..
    \label{eq:Aq}
\end{equation}
In this equation, the discrete sum over the various modes $n$, with $\hat{a}_n$ and $\hat{a}_n^\dagger$ representing the annihilation and creation operators, respectively, each mode is characterized by an eigenfrequency $\omega_{\rm rf}^n$ and a wavefunction $\vec{\epsilon}_n(\vec{r}\,)$. With a perfect conductor as the boundary, each resonant mode arises from solving the vacuum Maxwell equations $\Box \vec{A} = 0$, subject to the boundary conditions:
\begin{align}
    \hat{s} \times \vec{\epsilon}_n |_s = 0, \qquad
    \hat{s} \cdot \left(\vec{\nabla} \times \vec{\epsilon}_n\right)|_s = 0.
\end{align}
Here, the subscript $s$ denotes the surface of the cavity, with $\hat{s}$ as the normal vector. Moreover, the wavefunctions must satisfy the orthonormality condition:
\be \int_V \vec{\epsilon}_m^{\ *} \, \vec{\epsilon}_n\,  \D^3 \vec{r}  = \delta_{mn}.\label{norm} \ee
Using Eq.~\ref{eq:Aq}, the free Hamiltonian for the modes within the cavity is given by:
\be\begin{split}
    H_0 =  \frac{1}{2} \int_\text{V} \left(\vec{E}^2+\vec{B}^2\right)\, \D^3 \vec{r} = \sum_n \omega_{\rm rf}^n \left(\hat{a}_n^\dagger \hat{a}_n+\frac{1}{2}\right).\label{eq:Hcav}
    \end{split}
\ee
where $\vec{E} = -\partial \vec{A} / \partial t$ and $\vec{B} = \vec{\nabla} \times \vec{A}$ represent the electric and magnetic fields of the cavity modes, respectively.

We proceed by examining the interaction Hamiltonian concerning the spatial component of the effective current $\vec{J}_{\rm eff}$. A linear coupling with the vector potential $\vec{A}$ is established as follows:
\be\begin{split} H_{\text{int}} =&
    \int_\text{V}\,\vec{A}\cdot \vec{J}_{\rm eff}\, \D^3 \vec{r} \\
    =& \sum_n \sqrt{\frac{V}{2 \omega_{\rm rf}^n}}\, \eta_{n}\, \bar{J}_{\rm eff}\, \hat{a}_n^{\dagger}\, e^{-\mi \omega_{\rm rf}^n\, t} + h.c. .\label{cavalp}
  \end{split}
\ee
Here, $\eta_n$ represents the geometric overlap function between $\vec{\epsilon}_n$ and $\vec{J}_{\rm eff}$, defined as:
\begin{equation}
    \eta_{n} \equiv \frac{{\int}_{V} \, \vec{\epsilon}_{n} \cdot \vec{J}_{\rm eff}\,  \D^3 \vec{r}}{\sqrt{\int_{V}|\vec{J}_{\rm eff}|^{2}\,  \D^3 \vec{r}}},
    \label{etan}
\end{equation}
and $\bar{J}_{\rm eff}$ denotes the average current density within the cavity:
\begin{equation}
   \bar{J}_{\rm eff} \equiv \sqrt{ \frac{1}{V}\int_{V}|\vec{J}_{\rm eff}|^{2} \,  \D^3 \vec{r}}.\label{jeffint}
\end{equation}
On the other hand, the time component of $J^\mu_{\rm eff}$, corresponding to the effective charge density $\rho_{\rm eff}$, exclusively excites the irrotational mode, which does not experience resonant enhancement within the cavity context~\cite{hill2009electromagnetic}.

In addition to electromagnetic interactions involving the effective current, a cavity mode can also be excited by an external force applied to the cavity when a background mode is present. This force induces a displacement of the cavity walls, thereby instigating a power transition between the background and signal modes. This phenomenon can be characterized by a linear coupling resembling the expression in Eq.~(\ref{cavalp}), which will be demonstrated in Sec.~\ref{sec:Mechanical-deformation}.

\subsection{LC Circuit}
The resonant frequency of a cavity is intrinsically linked to its spatial dimensions, which poses a challenge in achieving a wide frequency tuning range outside the GHz regime. This limitation can be mitigated through the utilization of an LC circuit, which allows for precise tuning of the resonant frequency across different orders of magnitude below the GHz range. The LC circuit, comprising an inductor $L$ and a capacitor $C$, forms a resonant system with a resonant frequency given by $\omega_{\rm rf} = 1/\sqrt{LC}$. The system's free Hamiltonian, encompassing the energy stored within the inductor and the capacitor, is represented as follows:
\be     H_0 = \frac{\Phi^2}{2L} +\frac{Q^2}{2C} = \omega_{\rm rf} \left(\hat{a}^\dagger \hat{a}+\frac{1}{2}\right).\label{eq:Hc}\ee
Here, $\Phi$ denotes the magnetic flux traversing the inductor, and $Q$ signifies the charge accumulated in the capacitor. These serve as the canonical coordinate and its conjugate momentum, respectively, and are expressed in terms of annihilation and creation operators.

By introducing a pick-up loop, the LC circuit becomes capable of capturing an external magnetic flux $\Phi_\Psi$ and resonantly responding when the frequency of $\Phi_\Psi$ closely aligns with $\omega_{\rm rf}$. The interaction Hamiltonian, arising from a shift of the canonical coordinate $\Phi \rightarrow \Phi + \Phi_\Psi$ within Eq.~(\ref{eq:Hc}), is presented as follows:
\begin{align}
    H_{\rm int} = \frac{\Phi\,\Phi_\Psi}{L} = \sqrt{\frac{\omega_{\rm rf}}{2L}}\Phi_{\Psi} \,\hat{a}_n^{\dagger}\, e^{-\mi \omega_{\rm rf}\, t} + h.c..
    \label{eq:Hl}
\end{align}

The magnetic flux in the pick-up loop originates solely from the spatial component of the effective current. The effective currents induce a magnetic field expressed as:
\begin{equation}
        \vec{B}_{\Psi}(\vec{r}\,) \approx \int \frac{\vec{J}_{\mathrm{eff}} (\vec{r}^{\,\prime}) \times\left(\vec{r}-\vec{r}^{\,\prime}\right)}{4 \pi\left|\vec{r}-\vec{r}^{\,\prime}\right|^{3}}
       \D^3 \vec{r}^{\,\prime},
\end{equation}
where we neglect the time derivative terms, assuming that the Compton wavelength of the source is much larger than the spatial scale of the detector. The existence of a conducting shield modifies the response magnetic field, which can be described as a geometric overlapping between the spatial distribution of $\vec{J}_{\rm eff}$ and all the eigenmodes of the shielding cavity~\cite{Chaudhuri:2014dla}. We define a dimensionless geometric overlap function $\eta \sim \mathcal{O}(0.1)$ so that $\Phi_{\Psi} = \eta V \bar{J}_{\rm eff}$, with $\bar{J}_{\rm eff}$ defined in Eq.~(\ref{jeffint}).

\subsection{Circuit Representation of Cavity Modes}
\label{sec:circuit-representation}
Encapsulating an antenna within a cavity enables the investigation of specific cavity modes by analyzing the nodal flow across the antenna, expressed as:
  \be\begin{split}
        \Phi = \int_{\text{Ant}} \vec{A} (\vec{r},t) \cdot \D \vec{l}= \sum_n \kappa_n \hat{a}_n^\dagger e^{-\mi \omega_{\rm rf}^n t} + h.c..
        \label{def:Phi}
  \end{split}
\ee
Here, the integration spans the spatial extent of the antenna. The zero-point uncertainty associated with a specific cavity mode is captured by $\kappa_n$, defined as:
        \begin{equation}
        \kappa_n  \equiv \frac{1}{\sqrt{2 \omega_{\rm rf}^n}} \int_{\text{Ant}} \vec{\epsilon}_n(\vec{r}\,) \cdot \D \vec{l}.
        \label{eq: kappa}
    \end{equation}
Given the harmonic oscillator behavior of the Hamiltonians in Eqs.~(\ref{eq:Hcav}) and (\ref{eq:Hc}), it is feasible to model a cavity mode using an equivalent LC circuit. In this model, the nodal flux is defined as in Eq.~(\ref{def:Phi}), and the resonant frequency is given by $\omega_{\rm rf}^n = 1/\sqrt{L_n C_n}$. This analogy establishes a correspondence between the physical properties of cavity modes and circuit elements:
     \be
        L_n = \frac{2 \kappa_n^2}{\omega_{\rm rf}^n},\qquad
        C_n = \frac{1}{2 \kappa_n^2 \omega_{\rm rf}^n}.
    \label{eq: kappa_LC}
    \ee

\section{Axion, Dark Photon, and Gravitational Wave-Induced Signals}\label{sec:signals}

In this section, we explore the generation of effective current signals by axions, dark photons, and GWs. Additionally, we examine the mechanical vibrations induced by GWs, which can facilitate transitions between background and signal modes within cavities.

\subsection{Effective Currents}

The interaction Lagrangian between ultralight boson DM or HFGWs and electromagnetic fields can be reformulated as $A_\mu J^\mu_{\rm eff}$ to identify the effective currents $J^\mu_{\rm eff}$. An example of this is the axion-photon interaction, which is expressed as follows:
    \be \frac{1}{4} g_{a \gamma} a F_{\mu \nu} \tilde{F}^{\mu \nu} \rightarrow  J_{\text{eff}}^{a\,\mu} = g_{a\gamma} \tilde{F}^{\mu \nu} \partial_{\nu} a.
    \label{eq:aJ}\ee
Here, $g_{a \gamma}$ represents the axion-photon coupling constant, $a$ is the axion field, and $F_{\mu \nu}$ and $\tilde{F}^{\mu \nu}$ are the electromagnetic field strength tensor and its dual tensor, respectively. The Bianchi identity ensures that the derivative in Eq.~(\ref{eq:aJ}) affects only the axion field. In scenarios involving non-relativistic axion dark matter, the term with the time derivative primarily contributes to the axion's mass ($\omega_a \simeq m_a$), which is approximately $10^3$ times greater than that from the spatial derivative~\cite{Foster:2017hbq}. Consequently, a background magnetic field $\vec{B}_0$ is typically employed in electromagnetic resonant detection, leading to the spatial current:
    \be \vec{J}^a_{\rm eff} = g_{a \gamma} \omega_a a \vec{B}_0.\label{jaxion}\ee
Here, $\vec{B}_0$ may be static or oscillating, depending on the experimental setup. The latter corresponds to heterodyne upconversion, typically employing superconducting radio-frequency (SRF) cavities with significantly high quality factors ($Q_{\rm int} \gg 10^9$)~\cite{Berlin:2019ahk, Berlin:2020vrk, Berlin:2023grv}.

The subsequent example involves the kinetic mixing of the dark photon $A^{\prime \mu}$, whose interaction in the interaction basis results in:
    \be \epsilon m^2_{A^\prime} A^{\prime \mu} A_{\mu} \rightarrow J_{\text{eff}}^{A^\prime\,\mu} = \epsilon m^2_{A^\prime} A^{\prime \mu},\label{eq:JDP}\ee
Here, $\epsilon$ represents the kinetic mixing coefficient, and $m_{A^\prime}$ denotes the mass of the dark photon. Unlike the axion case, the effective current in Eq.~(\ref{eq:JDP}) depends solely on the dark photon field, thus an electromagnetic background field is not relevant. Specifically, the spatial component of the current is directly proportional to $\vec{A}^\prime$.

Finally, we consider the GW strain denoted as $h^{\mu\nu}$, which features a universal coupling to matter fields. Its interaction with electromagnetic fields yields:
\be  \frac{1}{2} h^{\mu\nu} T_{\mu\nu}^{\rm EM} \rightarrow J_{\text{eff}}^{h\,\mu} =  \partial_{\nu}\left(\frac{1}{2}hF^{\mu \nu} +{h^{\nu}}_{\rho}F^{\rho \mu}-{h^{\mu}}_{\rho}F^{\rho \nu}\right).\label{JGW} \ee
Here, $T_{\mu\nu}^{\rm EM}$ corresponds to the energy-momentum tensor of the electromagnetic field, and $h \equiv {h^{\rho}}_{\rho}$. Similar to the axion scenario, the generation of the effective current in Eq.~(\ref{JGW}) requires a background electromagnetic field, established within the proper detector frame. Notably, while the GW strain is often computed in the transverse-traceless (TT) frame characterized by the amplitude $h_0$, transitioning from the proper detector frame to $h_0$ introduces a scaling factor of $\omega_h^2 V^{2/3}$~\cite{Berlin:2021txa, Domcke:2022rgu}.

In scenarios characterized by a stationary magnetic field background, the derivative in Eq.~(\ref{JGW}) pertains exclusively to the GW strain~\cite{Berlin:2021txa}, resulting in:
    \begin{equation}
    \vec{J}^{\,h}_{\text{eff}} \simeq \omega_h^2 V^{\frac{1}{3}} B_0\, h_0\, \hat{j} (\vec{r}\,).\label{SCGW}
\end{equation}
where we assume $\omega_h \sim V^{-1/3}$. The spatially-dependent dimensionless vector $\hat{j} (\vec{r}\,)$ is related to the incoming direction and polarization of the GW. In the context of heterodyne upconversion detection, $\omega_{\rm rf}$ typically significantly outweighs $\omega_h$, consequently rendering the dominant component in Eq.~(\ref{JGW}) proportional to $\omega_{\rm rf} \sim V^{-1/3}$, thus leading to approximately the same expression as in Eq.~(\ref{SCGW}).

As elaborated in Sec.~\ref{sec:detectors}, the spatial component of the effective currents exhibits coupling with either a cavity mode or a circuit, expressed in the following Hamiltonian:
\begin{align}
    {H}_{\alpha}= \alpha \Psi \left(\hat{a}\, e^{\mi \omega_{\rm rf} t}+\hat{a}^\dagger\, e^{-\mi \omega_{\rm rf} t}\right)/\sqrt{2}.
\label{eq:hintA}
\end{align}
Here, the coupling coefficients $\alpha$ for the cavity and circuit scenarios are given by:
\begin{equation}
    \alpha_{\textrm{cav}}\Psi=\sqrt{\frac{V}{\omega_{\rm rf} }} \eta \bar{J}_{\rm eff}, \qquad   \alpha_{\textrm{LC}}\Psi=\sqrt{ \frac{\omega_{\rm rf}}{L}} \Phi_\Psi,
\end{equation}
respectively, derived in accordance with Eq.~(\ref{cavalp}) and Eq.~(\ref{eq:Hl}).

\begin{table*}[t]
\begin{center}
\begin{tabular}{|c||*{3}{c|}}\hline
\diaghead{\theadfont HaloscopeSource}%
  {Haloscope}{Source} & Axion & Dark Photon & HFGW\\\hline\hline
LC circuit & $g_{a\gamma}\eta B_0 V^{5/6} \omega_a \sqrt{\omega_{\text{rf}}}$ & $\epsilon \eta  V^{5/6} m_{A^\prime}^{2}  \sqrt{\omega_{\text{rf}}}$  &$ \eta B_0  V^{7/6}\omega^2_h \sqrt{\omega_{\text{rf}}}$ \\\hline
Cavity & $g_{a\gamma}\eta B_0 V^{1/2}  \omega_a /\sqrt{\omega_{\text{rf}}}$ & $\epsilon \eta  V^{1/2}m_{A^\prime}^{2} /\sqrt{\omega_{\text{rf}}}$ & $\eta B_0  V^{5/6}\omega_h^2/\sqrt{ \omega_{\text{rf}}}$ \\\hline
${\rm SRF^{EM}}$ &
$g_{a\gamma}\eta B_0  V^{1/2}\omega_a/ \sqrt{2 \omega_{\text{rf}}}$ & $\epsilon \eta  V^{1/2}m_{A^\prime}^{2} /\sqrt{\omega_{\text{rf}}}$ &$\eta B_0 V^{7/6}\omega_h^2\sqrt{\omega_{\text{rf}}/2}$\\\hline
${\rm SRF^{mech}}$ &
/ & / & $ \eta^t_p \eta_p^{h} L_p(\omega_h) B_0 V^{1/2} \omega_h^2 \sqrt{\omega_{\text{rf}}/2}$ \\\hline
\end{tabular}
\caption{Effective couplings $\alpha$, defined in Eq.~(\ref{eq:hintA}), between axions, dark photons, or HFGWs and sensor modes are listed for four detection schemes. Notably, ${\rm SRF^{EM}}$ for axions and HFGWs as well as ${\rm SRF^{mech}}$ refer to heterodyne upconversion using a background cavity mode.}
\label{table:alpha}
\end{center}
\end{table*}

The average current density, $\bar{J}_{\rm eff}$, can be directly inferred from Eq.~(\ref{jaxion}), Eq.~(\ref{eq:JDP}), and Eq.~(\ref{SCGW}) for the respective source. The magnetic field strength, $B_0$,  is defined as the average magnitude over both space and time within the relevant volume. The magnetic flux $\Phi_\Psi$, which traverses the pick-up loop of an LC circuit, is closely connected to the mean effective current density $\bar{J}_{\rm eff}$ by two factors: the spatial scale $V$ and a dimensionless geometric parameter $\eta$, resulting in $\Phi_\Psi \sim \eta V \bar{J}_{\rm eff}$. For simplicity, the inductance $L$ is assumed to be $V^{1/3}$. It is noteworthy that achieving optimal scaling for HFGW detection using an LC circuit necessitates a specialized design of the pick-up loops, as introduced in Ref.~\cite{Domcke:2022rgu}. The values of the parameter $\alpha$ for each source and detection scheme are detailed in Table.~\ref{table:alpha}. Here, SRF$^{\rm EM}$ for axions and HFGWs refers to heterodyne upconversion using a background cavity mode. This setup differs from the cavity employing a static magnetic field background by a factor of $\sqrt{2}$, due to signal modes being created at two sidebands around the pump mode and the time-averaged magnitude being $\sqrt{2}$ smaller than the oscillation amplitude.

\subsection{Mechanical Deformation}
\label{sec:Mechanical-deformation}
In addition to the microscopic interaction between GWs and electromagnetic fields, an alternative mechanism for inducing cavity mode transitions involves the deformation of the cavity's inner surface driven by GWs, as explored in Ref.~\cite{Berlin:2023grv} for the SRF cavity configuration. This deformation is characterized by local displacements from the equilibrium position, represented as:
\be \vec{U}(\vec{r}, t)= \sum_p x_p(t) \vec{U}_p(\vec{r}\,),\ee
where the subscript $p$ corresponds to a mechanical normal mode with a time-dependent mode function $x_p(t)$, and $\vec{r}$ denotes a point within the cavity shell with volume $V_S$. The function $\vec{U}_p(\vec{r}\,)$ describes the dimensionless spatial profile of the $p$-mode, satisfying the normalization condition:
\be
   \int_{V_S} \,  \vec{U}^*_p(\vec{r}\,) \cdot \vec{U}_q(\vec{r}\,)\,\rho_S (\vec{r}\,)\, \D^3 \vec{r}=\delta_{pq}\,M_S,
\ee
where $\rho_S(\vec{r}\,)$ and $M_S$ denote the mass density and total mass of the cavity shell, respectively. These mechanical normal modes can be excited by an external force, described by the following equation of motion:
\be
    \ddot{x}_p+\frac{\omega_p}{Q_p} \dot{x}_p+\omega_p^2 x_p =\frac{1}{M_S} \int_{V_S} \, \vec{f}(\vec{r}, t) \cdot \vec{U}^*_p(\vec{r\,})\,\D^3 \vec{r} \equiv \frac{F_p}{M_S}. \label{ftou}
\ee
Here, $\omega_p$ and $Q_p$ represent the resonant frequency and mechanical quality factor of the $p$-mode, respectively. The force density is denoted as $\vec{f}(\vec{r}, t)$, and $F_p$ represents the total force applied to the $p$-mode. The solution to this equation in the frequency domain is straightforwardly obtained as $x_p(\omega)=L_p(\omega) F_p(\omega) / M_S$, with $L_p(\omega)$ being the response function given by $(\omega^2 - \omega_p^2 + \mi \omega \omega_p/Q_p)^{-1}$.

GW strain acts as a tidal force on the cavity shell, exerting a force density $f_i = -\rho_S R_{i 0 j 0} r^j$ in the long-wavelength regime, where $r_j$ is defined in a coordinate system originating at the center of mass of the cavity. The Riemann curvature tensor is typically expressed in terms of the strain in the TT gauge, $R_{i 0 j 0} = -\ddot{h}_{ij}^{\mathrm{TT}} / 2 + \mathcal{O}(h^2)$. The strain-induced force on the $p$-mode, labeled as $F_p^h$, is then characterized through an overlapping function $\eta_p^h$ between the strain polarization basis $H_{ij}^{\mathrm{TT}}$ and the mechanical mode profile:
\be 
\begin{aligned}
    F^h_p(\omega)&=M_S V^{1/3} \eta^{h}_p \omega^2 h_0(\omega),\\
\eta_p^{h} &\equiv \frac{{H}_{i j}^{\mathrm{TT}}}{2V^{1 / 3} M_S} \left\vert \int_{V_{S}}\, \rho(\vec{r}\,)\,U_p^{i*}(\vec{r}\,)\, r^j\, \D^3 \vec{r} \,\right\vert.
\end{aligned}
\ee
Here, $h_0(\omega)$ represents the strain amplitude, satisfying the relation $h_{ij}^{\mathrm{TT}} \equiv h_0 H_{ij}^{\mathrm{TT}}$.

The deformation of the inner cavity surface leads to deviations from the orthonormality condition of the cavity modes in Eq.~(\ref{norm}), enabling the transition from the pump mode to the signal mode. The effective coupling between these modes in the frequency domain is identified as per the following equation~\cite{Bernard:2002ci, Berlin:2023grv}:
 \be \alpha\Psi(\omega) = \frac{\sqrt{2}}{2} \frac{ \omega_{\text{rf}}^{1/2}  B_0 V^{1/6}}{M_S} \sum_p \eta^t_p L_p(\omega-\omega_0) F_p^h(\omega-\omega_0).
    \label{HamiMech} \ee
Here, $\eta_p^t$ represents the transition form factor given by
\be     \eta^t_p=V^{1/3} \int_S \, \left( \vec{\epsilon}_0 \cdot \vec{\epsilon}_1{}^* - \frac{1}{\omega_{\text{rf}}^2} \left( \bm{\mathrm{curl}} \vec{\epsilon}_0 \right) \cdot \left( \bm{\mathrm{curl}} \vec{\epsilon}_1{}^* \right) \right)\,\D\vec{s} \cdot \vec{U}_p , \ee
where the integral is performed over the inner surface of the cavity. The subscripts $0$ and $1$ denote the pump and signal modes, respectively. For simplicity, it is assumed that the dominant contribution from GW is manifested in a single quadrupolar mechanical mode possessing the lowest $\omega_p$, as $\eta_p^h \propto 1/\omega_p^2$. The expression of the effective coupling is listed in Table~\ref{table:alpha} as well.

\section{Quantum Limit for Single-Mode Resonators}\label{sec:single-mode}
This section begins with a brief introduction to the input-output formalism used to derive the scattering matrix elements of single-mode resonators. From these elements, one can calculate the signal and noise power spectral densities (PSDs). We then explore how to optimize the scan rate by adjusting the readout coupling and distributing the integration time within the $e$-fold time. Additionally, we discuss how the optimized scan rate translates into the physics reaches of various detection schemes for the three sources outlined in Sec.~\ref{sec:signals}.

\subsection{Input-Output Formalism}
The Hamiltonian governing a resonant mode $\hat{a}$ is given by
\begin{equation}
H = H_{\rm free} + H_{\textrm{int}},
\end{equation}
where $H_{\rm free} = \omega_{\rm rf} (\hat{a}^\dagger \hat{a} + 1/2)$ represents the free Hamiltonian, and $H_{\textrm{int}}$ includes the interaction terms. Utilizing the Heisenberg equation in the interaction picture,
\begin{equation}
    \frac{\D}{\D t}\hat{a}=-\mi \Big[ \hat{a},H_\textrm{int} \Big],\label{eq:HEI}
\end{equation}
leads to the derivation of the system’s dynamics.

Before exploring specific interaction forms, it is essential to consider environmental factors that become relevant when the system is coupled to an external port, as described by the quantum Langevin equation:
\begin{equation}
  -\mi \Big[ \hat{a}, H_\textrm{env}^\textrm{p} \Big] = -\gamma_{\rm p} \hat{a} + \sqrt{2 \gamma_{\rm p}} \hat{u}_{\rm p},\label{eq:QLE}
\end{equation}
where $H_{\textrm{env}}^{\textrm{p}}$ represents the interaction with the environment via port p, $\gamma_{\rm p}$ denotes the dissipation coefficient, and $\hat{u}_{\rm p}$ corresponds to incoming noise as dictated by the fluctuation-dissipation theorem. The input-output relation is then derived from the port boundary condition:
\begin{equation}
    \hat{v}_{\rm p} = \hat{u}_{\rm p}- \sqrt{2 \gamma_{\rm p}} \hat{a},\label{eq:IOR}
\end{equation}
where $\hat{v}_{\rm p}$ designates the outgoing mode.

For a single-mode resonant detection system, two noise-contributing ports are relevant: intrinsic dissipation and readout. Together with the interaction with a potential signal, $H_{\alpha}$ introduced in Eq.~(\ref{eq:hintA}), the interaction Hamiltonian becomes:
\begin{equation}
    H_\textrm{int}=H_{\alpha}+H_\text{env}^{\rm \gamma}+H_\textrm{env}^{\gamma_r}.
    \label{eq: Hint}
\end{equation}
Utilizing Eq.~(\ref{eq:HEI}) and Eq.~(\ref{eq:QLE}), the equation of motion for $\hat{a}$ in the frequency domain can be solved as follows:
\begin{equation}
    \hat{a} = \frac{-\mi\alpha\Psi e^{-\mi \omega_{\rm rf} t}/\sqrt{2} + \sqrt{2\gamma} \hat{u}_a+\sqrt{2\gamma_r}\hat{u}_r}{\gamma +\gamma_r - \mi \Omega}, 
\end{equation}
where $\Omega \equiv \omega - \omega_{\rm rf}$ represents the frequency shift in the interaction picture, and $\gamma$ and $\gamma_r$ denote the dissipation coefficients of the two ports, with corresponding incoming modes $\hat{u}_a$ and $\hat{u}_r$, respectively. The intrinsic dissipation can alternatively be expressed in terms of the quality factor $Q_{\rm int}$, which is related to $\gamma$ by $Q_{\rm int} \equiv \omega_{\rm rf}/(2\gamma)$.

The observable is obtained from the readout port's outgoing mode, as per the input-output relation in Eq.~(\ref{eq:IOR}):
\begin{equation}
    \hat{v}_r=S_{0r} \left(\hat{u}_a - \frac{\mi \alpha e^{-\mi \omega_{\rm rf} t}}{2\sqrt{\gamma}} \Psi\right)+S_{rr} \hat{u}_r,
\end{equation}
The two scattering matrix elements, characterizing the propagation from the input to the output of different ports, are given by:
\be
     S_{0r} = -\frac{2\sqrt{\gamma\gamma_r}}{\gamma+\gamma_r-\mi\Omega}, \qquad
    S_{rr} = \frac{\gamma-\gamma_r-\mi\Omega}{\gamma+\gamma_r-\mi\Omega}.\label{eq:SM}
\ee
Here, the subscript $0$ represents the probing sensor, such as a cavity or a circuit discussed in Sec.~\ref{sec:signals}, and $r$ indicates the readout port.

When the integration time significantly exceeds the signal's correlation time, the PSD of the outgoing mode is typically taken to be the observable, defined by:
\begin{equation}
    S_{v_r}(\Omega)\delta(\Omega-\Omega^{\prime}) \equiv \braket{\hat{v}_r(\Omega)\hat{v}_r^*(\Omega^{\prime})},
    \label{eq:Svr}
\end{equation}
incorporating PSDs of both the signal and noise.

\subsection{Physics Reach}

According to Eq.~(\ref{eq: Hint}), apart from the signal port $H_\alpha$, fluctuations are induced by intrinsic dissipation and the readout port. The readout PSD in Eq.~(\ref{eq:Svr}), expressed as $S_{v_r}(\Omega) = S_{\text{sig}} + S_{\text{noise}}$, can be separated as:
\be\begin{split}
    S_{\text{sig}}  &= \left|S_{0r}\right|^2 \frac{\alpha^2}{4\gamma} S_\Psi,
    \\
    S_{\text{noise}}  &= \left|S_{0r}\right|^2 n_\occ+\left|S_{rr}\right|^2\frac{1}{2} + \frac{1}{2}.\label{eq:SN}
    \end{split}
\ee
Here, $S_\Psi$ represents the source PSD, which is related to its energy density $\rho_\Psi$ as follows:
    \begin{equation}
    S_\Psi = \frac{2\pi}{\omega^2}\frac{\D\rho_{\Psi}}{\D\omega}.
\end{equation} 
The three noise terms in Eq.~(\ref{eq:SN}) represent intrinsic fluctuation noise, the readout port noise, and additional amplifier noise, respectively. $n_{\text{occ}}$ represents the intrinsic noise occupation number, for example, the thermal fluctuation of a cavity mode gives $n_{\text{occ}}^{\rm th} =  1/2 + 1/(e^{\omega/T}-1)$, where $\omega$ is the frequency and $T$ is the temperature. In the zero-temperature limit, the sum of the noise precisely equals one due to unitarity, leading to the standard quantum limit of single-mode resonant detection~\cite{Chaudhuri:2018rqn, Chaudhuri:2019ntz}.

The sensitivity reach of each scan can be estimated by requiring that the signal-to-noise ratio (SNR) be of order one~\cite{Chaudhuri:2018rqn,Chaudhuri:2019ntz,Berlin:2019ahk,Lasenby:2019prg,Lasenby:2019hfz,Chen:2021bgy}, as described by the Dicke radiometer equation~\cite{Dicke:1946glx}:
\be \text{SNR}^2 = \frac{t_{\text{int}}}{2\pi} \int_0^\infty \left( \frac{S_{\text{sig}}}{S_\text{noise}} \right)^2 d\omega. \label{eq:SNR}\ee
Here, $t_{\text{int}}$ represents the integration time. The integrand in Eq.~(\ref{eq:SNR}) is decomposed into the product of two distributions: the signal PSD distribution $\alpha^4 S_\Psi^2/\gamma^2$ and the sensitive response function of the detector, characterized by $(|S_{0r}|^2 n_{\text{occ}}/S_{\text{noise}})^2$, which quantifies the ratio of intrinsic fluctuation noise against the total noise.

For convenience, we introduce the concept of the average source frequency as:
    \begin{equation}
    \overline{\omega}_{\Psi} \equiv \frac{\int\omega^2  S_\Psi\,\alpha^2/\gamma \, d\omega}{\int \omega S_\Psi\, \alpha^2/\gamma \, d\omega},
\end{equation}
and define the source bandwidth as:
    \begin{equation}
    \Delta\omega_{\Psi} \equiv 
    \frac{\int\omega^2  S_\Psi\,\alpha^2/\gamma \, d\omega}{\overline{\omega}_{\Psi}^2 S_{\Psi}(\overline{\omega}_{\Psi}) \alpha^2(\overline{\omega}_{\Psi})/\gamma(\overline{\omega}_{\Psi})}.
\end{equation}
The quality factor of the source is defined as $Q_\Psi \equiv \overline{\omega}_\Psi / \Delta\omega_{\Psi}$, which is $10^6$ for non-relativistic DM. For HFGWs, we parameterize the PSD as $S_h(\omega)=\Theta\left(\Delta\omega_h / 2-\left|\omega-\omega_h\right|\right) {h_0^2}/{\Delta\omega_h}$, where $\Theta$ is the Heaviside function and $\Delta\omega_h$ denotes the GW bandwidth.

Due to the common factor $|S_{0r}|^4$ of the signal and intrinsic fluctuation in Eq.~(\ref{eq:SNR}), the detector's effective sensitive width is approximately
\be \Delta \omega_r \equiv \int_0^\infty \left( \frac{\left|S_{0r}\right|^2 n_\occ}{S_{\text{noise}}}\right)^2 d\omega,\label{eq:omegar}\ee
within which the sensitivity to a source remains approximately the same order. 

The integral width in Eq.~(\ref{eq:SNR}) is determined by the minimum of $\Delta \omega_\Psi$ and $\Delta \omega_r$. Moreover, their maximum controls how the integration time $t_{\text{int}}$ of each scan is distributed within the total amount of time $t_e$ spent covering each $e$-fold of $\overline{\omega}_\Psi$:
\be t_\text{int}\simeq t_e\, \textrm{max} \left[\Delta \omega_\Psi, \Delta \omega_r\right]/\overline{\omega}_\Psi.\label{eq:tint}\ee
By considering only the dominant intrinsic noise within $\Delta \omega_r$, SNR$^2$ for a given hypothesis of $\overline{\omega}_\Psi$ is simplified to
\begin{equation}
    \text{SNR}^2 (\overline{\omega}_\Psi) \simeq \frac{t_e}{\overline{\omega}_{\Psi}}\Delta \omega_\Psi \Delta \omega_r \left.\frac{\alpha^4 S^2_{\Psi}}{32 \pi \gamma^2 n_{\occ}^2} \right|_{\omega = \overline{\omega}_{\Psi}}.\label{snrS}
\end{equation}

\begin{figure}[t]
    \centering
    \includegraphics[width=0.46\textwidth]{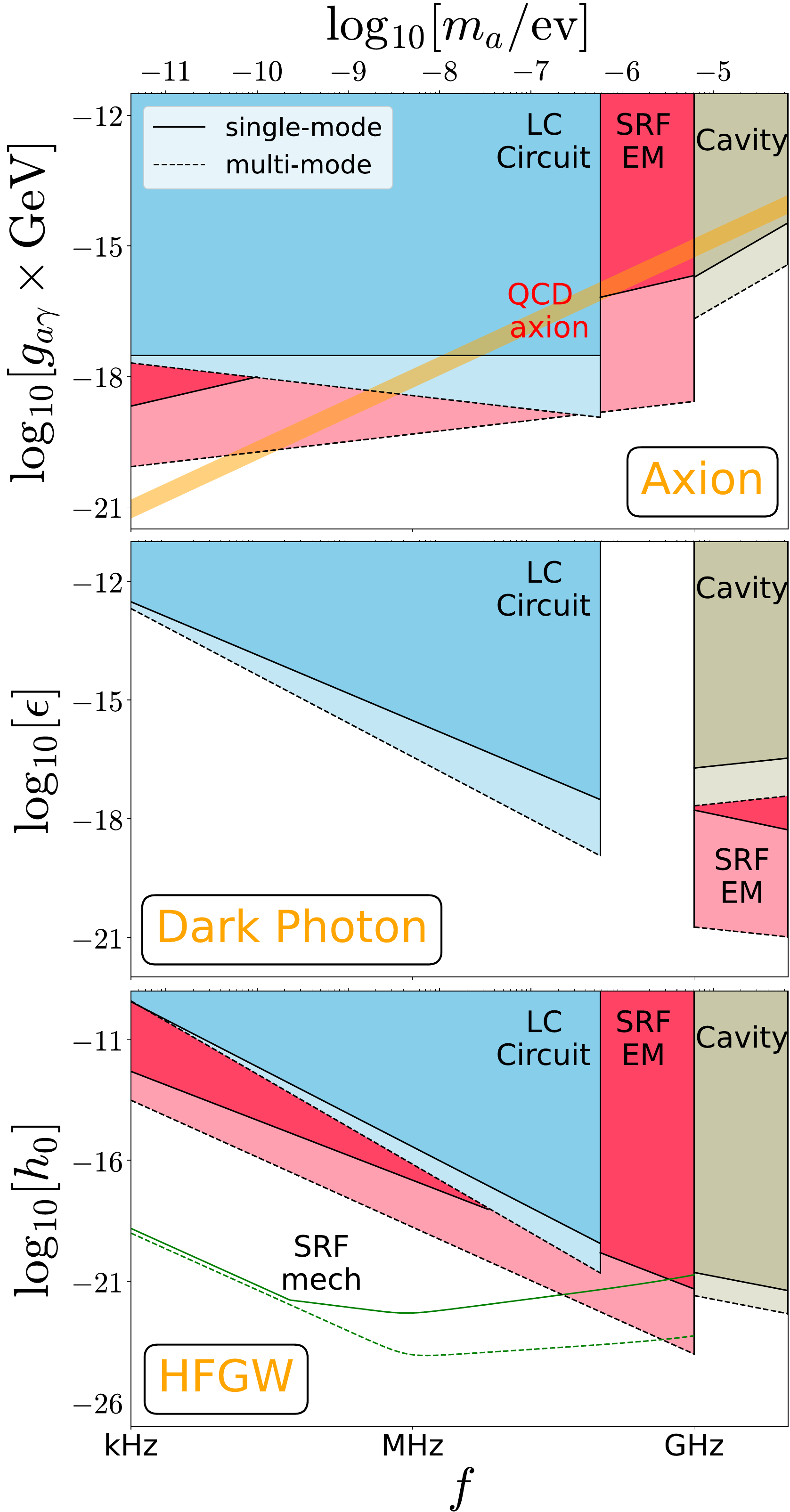}
    \caption{Sensitivity reach of axion and dark photon DM, and HFGW, depicted with solid lines for single-mode and dashed lines for multi-mode detection limits. The sensitivity thresholds for single-mode are based on requiring Eqs.~(\ref{eq:SNRaSM}, \ref{eq:SNRDPSM}, \ref{SNRL}) to reach $1$, while for multi-mode, the thresholds are derived from Eqs.~(\ref{eq:SNRaNN}, \ref{SNRDPMM}, \ref{eq:SNRHFGWMM}, \ref{eq:SRFEMA}, \ref{SNRMM}). The integration time allocated for each $e$-fold of the target source frequency is $t_e = 10^7$\,s. Benchmark parameters for different experimental setups are discussed in Sec.~\ref{sec:single-mode}.
    }
    \label{fig:bound}
\end{figure}

From the expression in Eq.~(\ref{snrS}), the key factor for a scan search is the response width $\Delta \omega_r$, defined in Eq.~(\ref{eq:omegar}), which is proportional to the scan rate as discussed in Refs.~\cite{Zheng:2016qjv, Malnou:2018dxn, HAYSTAC:2020kwv, Lehnert:2021gbj}. This parameter can be optimized by adjusting the readout coupling. Specifically, setting $\gamma_r = 2\gamma$ in the zero-temperature limit~\cite{Krauss:1985ub}, or $\gamma_r \simeq 2n_{\text{occ}}\gamma$ when $n_{\text{occ}} \gg 1$~\cite{Chaudhuri:2018rqn, Chaudhuri:2019ntz, Berlin:2019ahk, Lasenby:2019prg, Lasenby:2019hfz}, results in $\Delta\omega_r \simeq 3\gamma$ and $2n_{\text{occ}}\gamma$, respectively. Substituting these values back into Eq.~(\ref{snrS}) yields the sensitivity limit of single-mode resonators.

By incorporating the optimized $\gamma_r$ and the couplings detailed in Table~\ref{table:alpha} into Eq.~(\ref{snrS}), the resulting SNR$^2$ can be calculated for each instance of single-mode resonant detection, as well as for the three target sources:
\\
$\bullet$ Axion:
      \be\begin{split}
 {\rm LC\ Circuit}& :
   \frac{\pi}{2} g_{a\gamma}^4 \rho_{\text{DM}}^2 \eta^4 B_0^4 V^{{10}/{3}} Q_a Q_{\text{int}} t_e/T,\\
 {\rm Cavity}& :
 3 \pi \frac{g_{a\gamma}^4}{m_a^5} \rho_{\text{DM}}^2 \eta^4 B_0^4  V^2 Q_a Q_{\text{int}} t_e,\\
 {\rm SRF^{EM}}& :\frac{\pi}{8}\frac{g_{a\gamma}^4}{m_a^2} \rho_{\text{DM}}^2  \eta^4 B_0^4V^2  Q_a Q_{\text{int}} t_e/ (\omega_{\rm rf}^2T).
 \label{eq:SNRaSM}
     \end{split}\ee
     $\bullet$ Dark Photon:
\begin{align}
  {\rm LC\ Circuit}& :\frac{\pi}{2} \epsilon^4 m_{A^\prime}^{4} \rho_{\text{DM}}^2 \eta^4  V^{{10}/{3}} Q_{A^\prime} Q_{\text{int}} t_e/T,\nn\\
 {\rm Cavity}& :3 \pi \frac{\epsilon^4}{m_{A^\prime}} \rho_{\text{DM}}^2 \eta^4  V^2Q_{A^\prime} Q_{\text{int}} t_e, \label{eq:SNRDPSM} \\
   {\rm SRF^{EM}}& :\frac{\pi}{2} \epsilon^4m_{A^\prime}^2 \rho_{\text{DM}}^2 \eta^4  V^2 Q_{A^\prime} Q_{\text{int}} t_e/(\omega_{\rm rf}^2T).\nn  
\end{align}
       $\bullet$ HFGW:
\begin{align}
 {\rm LC\ Circuit}& :\frac{1}{8\pi} h_0^4 \omega_h^8\eta^4  B_0^4 V^{{14}/{3}}  Q_h Q_{\text{int}} t_e/T,\nn\\
{\rm Cavity}& :\frac{3}{4\pi} h_0^4 \omega_h^{3} \eta^4 B_0^4 V^{{10}/{3}}   Q_{h}Q_{\text{int}}t_e,\nn\\
 {\rm SRF^{EM}}& :   \frac{1}{32\pi} h_0^4 \omega_h^6 \eta^4  B_0^4 V^{{14}/{3}}  Q_h Q_{\rm int} t_e \omega_{\rm rf}^2/T,\nn\\
 {\rm SRF^{mech}}& : \frac{h_0^4 \omega_h^6 \vert \eta_p^t\,\eta_p^h\,L_p(\omega_h)\vert^4
 B_0^4 V^2 Q_h Q_{\text{int}} t_e \omega_{\text{rf}}^2}{16\pi T N_M}\nn \\
 &\quad \times {\rm min}\left(1, \frac{Q_{\rm int}\omega_h}{T N_M}\right).
 \label{SNRL}
\end{align}
In these expressions, we assume the thermal occupation number for the cavity to be $n^{\rm th}_{\rm occ} = 1/2$, and for the LC circuit and SRF$^{{\rm EM}/{\rm mech}}$, $n^{\rm th}_{\rm occ} = T/\omega$, which applies respectively when $T \ll \omega$ or $T \gg \omega$. For mechanical resonance, the SRF cavity is designed to maximize mechanical vibrations, which may cause the thermal vibrations of the cavity shell to transition the pump mode at frequency $\omega_0$ into the signal mode. This mechanism significantly contributes to $n_{\text{occ}}$, surpassing the intrinsic cavity mode thermal noise at low frequencies, expressed as~\cite{Berlin:2023grv}:
\be     n_\occ^{\text{mech}} = \frac{Q_{\text{int}}}{4\pi} \frac{B_0^2 V^{1/3}}{M_S^{2} } \frac{\omega_0^5}{\omega^3\omega_{\text{rf}}^2} \sum_p  S_{F_p^T}  \vert \eta_p^t\, L_p(\omega-\omega_0) \vert^2.
    \label{mechnoise} \ee
Here, $F_p^T$ represents the thermal vibration force, whose PSD is given by $S_{F_p^T} = 4\pi M_S \omega_p T / Q_p$, according to the equipartition theorem. In Eq.~(\ref{SNRL}), a dimensionless function $N_M(\omega_h)$ is defined as:
\be
    N_M(\omega_h)\equiv 1 + Q_{\text{int}} \,\omega_{\text{rf}}\, B_0^2\,S_{F_p^T} \vert\eta_p^t\, L_p(\omega_h) \vert^2\, V^{1/3}/(\pi T M_S^2),
\ee
where the first term corresponds to the intrinsic thermal noise of the cavity modes, and the second term is the noise induced by mechanical resonance. At frequencies below $\mathcal{O}(100)$\,kHz, the total contribution to $n_{\rm occ} = N_M n^{\rm th}_{\rm occ}$ increases significantly, causing the optimized bandwidth $\Delta \omega_r$ to exceed $\omega_h$. To address this, we impose a cutoff such that $t_{\text{int}}$ remains below $t_e$, achieved by applying the min-function at the end of Eq.~(\ref{SNRL}). We further assume that $\omega - \omega_0 \approx m_a$ or $\omega_h$, and that $\omega \approx \omega_0 \approx \omega_{\rm rf} \gg m_a$ or $\omega_h$ for setups involving heterodyne upconversion. For setups without heterodyne upconversion, including dark photon searches with SRF$^{\rm EM}$, we assume $\omega_{\rm rf} \approx \omega \approx m_a$, $m_{A^\prime}$, or $\omega_h$.

Requiring each case in Eqs.~(\ref{eq:SNRaSM},\ref{eq:SNRDPSM},\ref{SNRL}) to reach $\mathcal{O}(1)$, we can establish the sensitivity reach for the single-mode resonant detection, as illustrated in solid lines in Fig.~\ref{fig:bound}. The $e$-fold time is set at $t_e = 10^7$\,s. The relevant benchmark parameters of experiments are as follows:\\
\\
$\bullet$ LC circuit: $B_0 = 4\, \textrm{T}$, $Q_{\text{int}} = 10^6$, $T = 0.01$\,K, $V = 1\,  {\rm m}^3$;\\
$\bullet$ Cavity: $B_0 = 4\, \textrm{T}$, $Q_{\rm int}=10^4$, $T = 0.01\,$K, $V = 1\,  {\rm m}^3$;\\
$\bullet$ SRF: $B_0 = 0.2\, \textrm{T}$, $Q_{\textrm{int}} = 10^{12}$, $T = 1.8$\,K, $V = 1\,  {\rm m}^3$, $\omega_{\rm rf} \approx \omega_0 = 2\pi\,\text{GHz}$. \label{SRF_para}\\
\\
We consider the frequency ranges for traditional and SRF cavities used in dark photon searches to be between GHz and $10$\,GHz, LC circuits between kHz and $100$\,MHz, and heterodyne upconversion involving the SRF cavity between kHz and GHz. The axion-photon coupling $g_{a\gamma}$ and the dark photon kinetic mixing coefficient $\epsilon$ are included in $\alpha$, while $S_\Psi$ incorporates $h_0^2$. For axion and dark photon dark matter, we set both $Q_a$ and $Q_{A^\prime}$ at $10^6$ and assume a local dark matter density of $\rho_{\text{DM}} = 0.45\,\textrm{GeV/cm}^3$. The HFGW spectrum is generally model-dependent; we assume a quality factor, $Q_h = 10^3$, aligned approximately with the predictions for primordial black hole binary inspiral, considering frequency shifts within $t_{\text{int}}$~\cite{Domcke:2022rgu, Gatti:2024mde}. This assumption can be generalized to other sources by scaling $Q_h$ according to Eq.~(\ref{SNRL}). The geometric overlapping factor $\eta$ is set to $1$ for the axion and $1/\sqrt{3}$ for the dark photon, reflecting the projections of their random polarizations onto an axis. For GW detection involving electromagnetic coupling, $\eta$ is set to $1/10$, as outlined in Refs.~\cite{Berlin:2021txa, Domcke:2022rgu}. In contrast, for mechanical coupling, a specialized cavity shape is designed to maximize deformation-induced transitions, with selected parameters including $Q_p=10^6$, $\omega_p=10$\,kHz, $M_S=10$\,kg, $\eta_p^t=1$, and $\eta_p^h=0.18$~\cite{Berlin:2023grv}.

\section{Broadened Response Width in Multi-Mode Resonators}\label{sec:multi-mode}
As discussed in Sec.~\ref{sec:single-mode}, the scan rate of a detector is reflected in the effective response width, defined in Eq.~(\ref{eq:omegar}) as the range where intrinsic fluctuation noise dominates over readout noise. One method to exceed the quantum limit for scan rates involves squeezing techniques that reduce readout noise levels~\cite{Zheng:2016qjv, Malnou:2018dxn, HAYSTAC:2020kwv, Lehnert:2021gbj}. However, because the single-mode resonant response is a Lorentzian function that decreases quadratically as $|\Omega|$ increases in off-resonant regions, significantly extending the response width is challenging.

\begin{figure*}[t]
    \centering
    \includegraphics[width=0.9\textwidth]{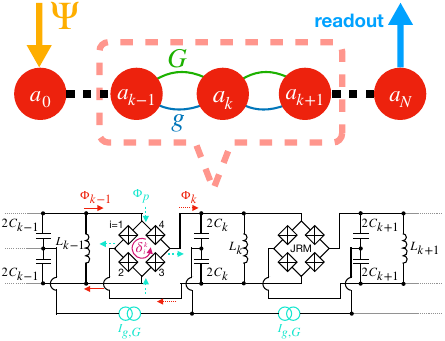}
    \caption{\textbf{Top:} Illustration of a chain of resonant modes as described by Eq.~(\ref{eq:lag_ch}). The lowest mode $a_0$ serves as the probing sensor, and the highest mode $a_N$ functions as the readout. Each pair of adjacent modes is interconnected through both beam-splitter-type (blue) and non-degenerate parametric (green) interactions. 
    \textbf{Bottom:} Circuit diagrams showing modes $a_{k-1}, a_k$, and $a_{k+1}$ connected via Josephson Ring Modulators (JRMs), detailed in Sec.~\ref{sec:JRMrealization}. Each JRM consists of four Josephson junctions (boxes), labeled $i = 1, 2, 3, 4$. The phases $\delta_i^k$ of these junctions are defined in a counter-clockwise direction within each JRM (purple).
    These JRMs facilitate three-wave mixing among $\Phi_{k-1}$ (solid red), $\Phi_k$ (dashed red), and the pumping mode $\Phi_p$ (dashed cyan), driven by a current source $I_{g, G}$ at the bottom.}
    \label{fig:ch}
\end{figure*}

Alternatively, by incorporating multiple auxiliary modes connected to the probing sensor, the direct response to signals can be significantly broadened and amplified without increasing the readout noise level~\cite{WLC, Chen:2021bgy, Wurtz:2021cnm}. For instance, consider the non-Hermitian chain structure modeled as a chain of modes with the interaction Hamiltonian~\cite{McDonald:2018cdg, McDonald:2020ywn}:
\begin{equation}
    H_{\rm ch}=\sum_{k=0}^{N-1}\left( \mi g \hat{a}_k \hat{a}^{\dagger}_{k+1} + \mi G \hat{a}_k \hat{a}_{k+1} +h.c. \right),
    \label{eq:lag_ch}
\end{equation}
where the parameters $g$ and $G$ represent the couplings for beam-splitter-type and non-degenerate parametric interactions, respectively~\cite{PCLC, Bergeal:2010, PhysRevB.87.014508, Chapman_2016, Frattini_2017}. This system comprises $N+1$ modes denoted by $\hat{a}_k$, with each adjacent pair linked by the two types of interactions. The dynamics described by Eq.~(\ref{eq:lag_ch}) can be interpreted as two copies of the Hatano-Nelson model, where two groups of quadratures are amplified in opposite directions~\cite{McDonald:2020ywn}. We designate $\hat{a}_0$ as the probing sensor, while the readout port is connected to the last mode $\hat{a}_N$, as illustrated in the top panel of Fig.~\ref{fig:ch}. The application of the $N = 1$ model of Eq.~(\ref{eq:lag_ch}) for axion DM was previously discussed in Refs.~\cite{Wurtz:2021cnm, Jiang:2022vpm}.

This section will focus on two examples of multi-mode systems, including the chain described in Eq.~(\ref{eq:lag_ch}), and a binary tree structure that amplifies both quadratures~\cite{Chen:2021bgy}. We will derive the corresponding input-output formalism and demonstrate that their response width can be significantly broadened to as large as the order of the resonant frequency.

\subsection{Scattering Matrices For Non-Hermitian Chain}
\label{sec:nhc}

We begin with a chain of resonant modes, where each neighboring pair is linked through both beam-splitter-type interactions and non-degenerate parametric interactions. The system is governed by the Hamiltonian:
      \be\begin{split}
    H_\text{ch}&=\sum_{k=0}^{N-1}\left( \mi  |g| e^{\mi \varphi^g_k} \hat{a}_k \hat{a}^{\dagger}_{k+1} + \mi |G| e^{\mi \varphi^G_k} \hat{a}_k \hat{a}_{k+1} +h.c. \right)\\
        &=\sum_{k=0}^{N-1} \left[(|g|-|G|) X_k \tilde{Y}_{k+1} - (|g|+|G|) \tilde{X}_{k+1} Y_k \right],\label{eqHch}
     \end{split}\ee
where $\varphi^{g/G}_k$ denote the relative phases of the couplings. Quadrature operators are introduced as follows:
\begin{equation}
\begin{aligned}
    &X_{k}\equiv\frac{e^{\mi \varphi^+_k} }{\sqrt{2}}  \hat{a}_k + h.c., \qquad Y_{k}\equiv \frac{e^{\mi \varphi^+_k} }{\sqrt{2  }\mi}  \hat{a}_k + h.c.,\\
    &\tilde{X}_{k}\equiv\frac{e^{\mi \varphi^-_{k-1}} }{\sqrt{2}}  \hat{a}_{k} + h.c., \quad \, \tilde{Y}_{k}\equiv \frac{e^{\mi \varphi^-_{k-1}} }{\sqrt{2} \mi}  \hat{a}_{k} + h.c.,\label{XY}
\end{aligned}    
\end{equation}
where $\varphi^\pm_k \equiv (\varphi^G_k \pm \varphi^g_k)/2$. As depicted in the top panel of Fig.~\ref{fig:ch}, the probe sensor mode and the readout port are strategically positioned at the two opposing termini of the chain, specifically labeled as $\hat{a}_0$ and $\hat{a}_N$, respectively.

Assuming zero relative phases $\varphi^{g/G}_k = 0$, we simplify the system's dynamics, leading to decoupled equations of motion for $Y_k = \tilde{Y}_{k}$:
\begin{equation}
\begin{aligned}
    &(\gamma- \mi \Omega)Y_0 + (|g|-|G|)Y_1=\sqrt{2 \gamma} \hat{u}_0,\\
    &(\gamma- \mi \Omega)Y_k + (|g|-|G|)Y_{k+1} - (|g|+|G|)Y_{k-1}=\sqrt{2 \gamma} \hat{u}_k, \\
    &(\gamma+\gamma_r- \mi \Omega)Y_N - (|g|+|G|)Y_{N-1}=\sqrt{2 \gamma} \hat{u}_N + \sqrt{2 \gamma_r} \hat{u}_r,
\end{aligned}
\label{eom_ch}
\end{equation}
where all intrinsic dissipation coefficients are simplified as $\gamma$. Recursive relations for $Y_k$ are derived from the first two lines of Eq.~(\ref{eom_ch}):
\begin{equation}
\begin{aligned}
    Y_k=y_k &\sum_{j=0}^k \prod_{m=j}^{k-1}  \mathcal{G}^{k-j}  y_m \sqrt{2 \gamma} \hat{u}_j -\mathcal{G} y_k  Y_{k+1}, \label{eqYk}
\end{aligned}
\end{equation}
where the series $y_k$, $\mathcal{J}$, and $\mathcal{G}$ are defined as:
\begin{equation}
\begin{aligned}
    &y_0\equiv{1}/{\left(\gamma-\mi \Omega\right)},\quad y_k\equiv{1}/\left(\gamma-\mi \Omega + \mathcal{J}^2 y_{k-1}\right),\\
    &\mathcal{J} \equiv (|g|^2-|G|^2)^{1/2}, \quad \mathcal{G} \equiv |g|+|G|.
\end{aligned}
\end{equation}
Consequently, the solution for the last mode $Y_N$ is expressed as:
\begin{equation}
\begin{aligned}
    Y_N=&\frac{\sqrt{2\gamma_r}\, \hat{u}_r+\sum_{k=0}^{N} \prod_{j=k}^{N-1} \,  \mathcal{G}^{N-k} y_j  \sqrt{2 \gamma}\hat{u}_k }{\gamma+\gamma_r-\mi\Omega+\mathcal{J}^2 y_{N-1}},
    \label{eq:YN}
\end{aligned}
\end{equation}
which provides the scattering matrix elements entering the readout port:
\be
    S_{kr}=\frac{-2\sqrt{\gamma \gamma_r} \mathcal{G}^{N-k}f_k}{\gamma_r f_N+f_{N+1}},\qquad
    S_{rr}=\frac{-\gamma_r f_N+f_{N+1}}{\gamma_r f_N+f_{N+1}},\label{SMEAC}
\ee
where $f_k$ is defined as:
\begin{equation}
    f_k \equiv \prod_{j=0}^{k-1}\frac{1}{y_j}=\sum_{j=0}^{\lfloor k/2 \rfloor}C_{k-j}^j (\gamma -\mi \Omega)^{k-2j} \mathcal{J}^{2j}.
    \label{eq: def_f}
\end{equation}
Here, $\lfloor \cdots \rfloor$ denotes the floor function that takes the integer part, and $C_{x-j}^j$ is the binomial coefficient. The stability condition $|g| > |G|$~\cite{McDonald:2018cdg} arises from the poles of the scattering matrix elements, assuming negligible $\gamma$.

Similarly, the equations of motion for $X_k = \tilde{X}_k$  differ from those for $Y_k$ only by the sign change in $|G|$, turning $|G| \rightarrow -|G|$. When the magnitude of $|g|$ significantly surpasses both $\gamma$ and $\mathcal{J}$, the progression of $Y_k$ or $X_k$ results in consecutive increments or decrements, respectively.

In the presence of nonvanishing relative phases $\varphi^{g/G}_k$, the amplification of the $k$-th mode to the $(k+1)$-th mode, denoted as $Y_k$, experiences a misalignment with the previously amplified $\tilde{Y}_k$. Instead, as defined earlier, they are related by a linear transformation:
\begin{equation}
    \tilde{X}_{k}=X_{k} \cos \theta_k - Y_{k} \sin \theta_k,\quad
    \tilde{Y}_{k}=X_{k} \sin \theta_k + Y_{k} \cos \theta_k.
    \label{eq: redef_XY}
\end{equation}
Here, $\theta_{k} \equiv (\varphi^G_{k-1} - \varphi^g_{k-1} - \varphi^G_{k} - \varphi^g_{k})/2$. Consequently, the amplification process experiences a sequence of attenuation factors expressed as the product $\Pi_{k=1}^{N-1} \cos \theta_k$. It is important to note that even with an initial calibration of all relative phases, fluctuations in the phases of the pumping modes can still contribute to $\varphi^{g/G}_k$~\cite{PhysRevB.96.104503}. Therefore, it is necessary to recalibrate the phase before each data acquisition cycle, as we discuss later.

\subsection{Scattering Matrices For Binary Tree}\label{sec:binarytree}
\begin{figure}[htb]
    \centering
    \includegraphics[width=0.45\textwidth]{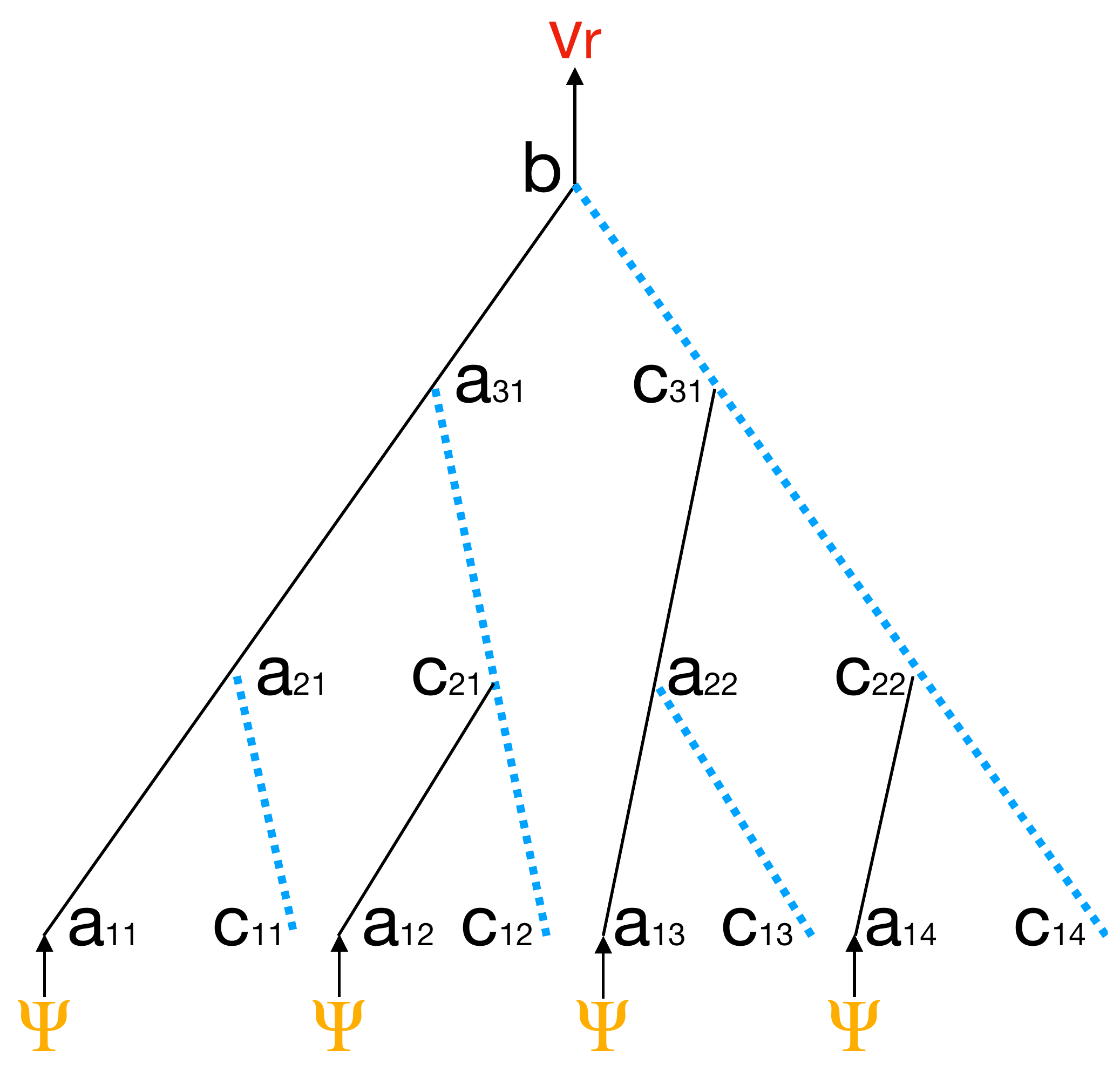}
    \caption{An illustration of a multi-mode resonator with a binary tree structure for $N = 3$. Black lines indicate beam-splitter-like interactions, and blue dashed lines represent non-degenerate parametric interactions. At the lowest level, a network of $2^N$ sensors can be positioned, facilitating the probing of $\Psi$. The remaining modes are auxiliary modes.}
    \label{fig:BT}
\end{figure}

We also consider a multi-mode resonator with a binary tree structure~\cite{Chen:2021bgy}. The corresponding Hamiltonian, denoted as $H_{\textrm{BT}}$, is formulated as:
\begin{equation}
\begin{aligned}
    H_{\textrm{BT}}=&\sum_{i=2}^N \sum_{j=1}^{2^{N-i}} \mi\,\hat{a}_{ij} \left( g \hat{a}_{i-1,2j-1}^{\dagger}+ G \hat{c}_{i-1,2j-1} \right)
    \\
    &+\sum_{i=2}^N \sum_{j=1}^{2^{N-i}} \mi\, \hat{c}_{ij} \left(  g \hat{a}_{i-1,2j}^{\dagger}+ G \hat{c}_{i-1,2j} \right)\\
    &+\mi\, \hat{b}\,\left( g \hat{a}_{N1}^{\dagger}+ G \hat{c}_{N1}\right)
    + h.c.,
\end{aligned}
\label{HBT}
\end{equation}
Similar to the chain-like structure in Eq.~(\ref{eqHch}), the binary tree encompasses both beam-splitter-type interactions and non-degenerate parametric interactions. However, it introduces auxiliary modes and connect every two adaject modes by either one coupling only. An illustrative example of the model with $N=3$ is presented in Fig.~\ref{fig:BT}.

The equation of motion for a mode situated within the middle of the network can be expressed as:
\begin{equation}
\begin{aligned}
    \hat{c}_{ij}=\,&\frac{\sqrt{2\gamma} \left[ \hat{u}_{ij}^c - \left(g^*  \hat{u}_{i-1,2j}^a + G^*  \hat{u}_{i-1,2j}^{c  \dagger}\right)/(\gamma - \mi \Omega)\right]}{\gamma-\mi \Omega +\mathcal{J}^2/(\gamma -\mi \Omega)}\\&+ \cdots,
\end{aligned}
\label{eomcij}
\end{equation}
where $\cdots$  accounts for terms involving modes not directly connected to $\hat{c}_{ij}$ as well as the coupling with the higher-level mode. The equations of motion for $\hat{a}_{ij}$ differ from Eq.~(\ref{eomcij}) only in the subscript, changing from $2j$ to $2j-1$. For $|g|\approx|G| \gg \gamma$, the propagation from $\hat{u}_{i-1,2j}^a$ or $\hat{u}_{i-1,2j}^{c \dagger}$ to $\hat{c}_{ij}$ in Eq.~(\ref{eomcij}) is amplified by a factor of $|g/(\gamma - \mi \Omega)|$, achieving significant amplification similar to the non-Hermitian chain model.

Notably, in the binary tree configuration, both quadratures of each mode within this structure are equally amplified, thus mitigating the issues of phase fluctuations encountered in the non-Hermitian chain~\cite{Chen:2021bgy}.

In this setup, approximately $2^N$ sensors can be positioned at the bottom level to probe $\Psi$, each undergoing successive amplification as they propagate toward $\hat{b}$. The coherent nature of the background source can be exploited to enhance sensitivity: signals from different probing sensors can be coherently combined at the readout port, boosting the signal PSD by a factor of $2^{2N}$. Although intrinsic thermal noise from each probing sensor also contributes, these fluctuations are uncorrelated across modes and thus grow only as $2^N$. This results in a net enhancement of the SNR$^2$ by a factor of $2^N$.

\subsection{Response Width in Multi-Mode Resonators}

As discussed in Sec.~\ref{sec:single-mode}, the response width determines the sensitivity reach of detectors. This width can be optimized numerically using Eq~(\ref{eq:omegar}) or approximated by identifying the range where intrinsic noise from $\hat{a}_0$ dominates over other noise contributions, as expressed in:
\begin{equation}
    \left|S_{0r}\right|^2 n_\occ \gtrsim \left|S_{rr}\right|^2\frac{1}{2} + \frac{1}{2} + \sum_{k=1}^{N}\left|S_{kr}\right|^2 n_\occ.
    \label{eq:bw_nhc}
\end{equation}
Here, the left-hand side represents intrinsic noise from $\hat{a}_0$, while the right-hand side accounts for readout noise and intrinsic noise from higher resonators.

The goal of optimization is to identify the best set of free parameters that maximizes the response width, as specified by Eq.~(\ref{eq:bw_nhc}), for a fixed number of chains, $N$. In principle, this condition depends on five parameters: $|g|$, $\gamma$, $n_\occ$, $\gamma_r$, and $\mathcal{J}$. We assume that $\gamma$ and $n_\occ$ are optimized experimentally to be as small as possible, and that $|g|$ is maximized to enhance $|S_{0r}|$. The remaining free parameters, $\gamma_r$ and $\mathcal{J}$, are tuned for optimization.

To begin, we examine the properties on the left-hand side of Eq.~(\ref{eq:bw_nhc}), particularly the scattering matrix $|S_{0r}|^2$. From Eqs.~(\ref{SMEAC}) and (\ref{eq: def_f}), we obtain
\begin{equation}
\begin{aligned}
    & |S_{0r}(\Omega = \Delta \omega_r)|^2 \\
    =&  \, \frac{4\gamma\gamma_r\mathcal{G}^{2N}|f_0|^2}{|\gamma_rf_N+f_{N+1}|^2}\\
     \approx & \, 4 \gamma\gamma_r\mathcal{G}^{2N} /  \left|\mathcal{J}^{N}\gamma_r \sum_{j=0}^{\lfloor N/2 \rfloor}C_{N-j}^j \left(-\mi \frac{\Delta\omega_r}{\mathcal{J}}\right)^{N-2j}\right.\\
    &\left.+\,\mathcal{J}^{N+1} \sum_{j=0}^{\lfloor (N+1)/2 \rfloor}C_{N+1-j}^j \left(-\mi \frac{\Delta\omega_r}{\mathcal{J}}\right)^{N+1-2j}\right|^2 \\
     \approx &  \,  \frac{4 \gamma\gamma_r\mathcal{G}^{2N}} { \Delta\omega_r^{2N}(\gamma_r+\Delta\omega_r)^2}.
\end{aligned}\label{eq:S_0r}
\end{equation}
In the third line, we neglect $\gamma$ in the expression for $f_N=\Sigma_{j=0}^{\lfloor N/2 \rfloor}C_{N-j}^j (\gamma -\mi \Omega)^{N-2j} \mathcal{J}^{2j}$, taking $f_0 = 1$, and evaluate the bandwidth at the boundary of the response width, $\Omega = \Delta \omega_r$. The last line is derived from the following three regions:
\begin{itemize}
    \item When $\mathcal{J} < \Delta \omega_r$, the highest-order polynomial term with $j = 0$ dominates in each sum, leading directly to the final expression.
    \item When $\mathcal{J} > \Delta \omega_r$, the condition is inconsistent. In this region, $|S_{0r}|^2$ remains nearly constant for $\Omega < \mathcal{J}$, dominated by the $0$-th order term in one of the polynomial series with either $j = \lfloor N/2 \rfloor$ or $\lfloor (N+1)/2 \rfloor$. $|S_{0r}|^2$ only begins to decrease significantly for $\Omega > \mathcal{J}$, ensuring that the bandwidth satisfies $\Delta \omega_r > \mathcal{J}$.
    \item When $\mathcal{J} \approx \Delta \omega_r$, both sums become approximately 1 based on numerical evaluation. Consequently, the term $|\cdots|^2$ can be approximated as $|\mathcal{J}^N \gamma_r + \mathcal{J}^{N+1}|^2$, which leads to the final expression by setting $\mathcal{J} \approx \Delta \omega_r$.
\end{itemize}

To proceed, we first determine the range of parameters that satisfies the necessary condition from Eq.~(\ref{eq:bw_nhc}), where the dominance of the left-hand side over the readout noise—characterized by a nearly constant PSD of approximately $1$ for sufficiently small $\gamma$—is required. This condition is expressed as
\begin{equation}
    |S_{0r}(\Omega = \Delta \omega_r)|^2n_{\rm occ}\approx\frac{4n_{\rm occ}\gamma\gamma_r\mathcal{G}^{2N}}{\Delta\omega_r^{2N}(\gamma_r+\Delta\omega_r)^2} \geq 1.
    \label{eq:C1}
\end{equation}
which must be satisfied to achieve an optimal response width. Maximizing $\Delta \omega_r$ for this relation is achieved by tuning $\gamma_r$, which leads to the simple conclusion that $\gamma_r \approx \Delta \omega_r$. It is important to note that this relation depends only marginally on the value of $\mathcal{J}$. As we will discuss later, variations in $\mathcal{J}$ affect the flatness of the PSD within $\Delta \omega_r$, since in the region where $\Omega \gg \mathcal{J}$, the PSD decreases rapidly.

Given this, we take $\mathcal{J} = \gamma_r = \Delta \omega_r$ as our benchmark parameters. This yields the expression $|S_{0r}|^2 \sim \gamma \mathcal{G}^{2N} /\Delta \omega_{r}^{2N+1}$, and the optimized response width, derived from Eq.~(\ref{eq:C1}), becomes
\be  \Delta \omega_{r}^{\rm opt} \simeq \Big( \gamma\, n_\occ\,  \mathcal{G}^{2N} \Big)^{1/(2N+1)}.\label{eq:rwo}\ee
In the limit of large $N$, this width converges to $\mathcal{G}$, provided that $\gamma n_{\rm occ}$ remains finite. Moreover, we can use $\mathcal{J} = \Delta \omega_r^{\rm opt}$ to solve for the exact values of $\mathcal{J}$ and $\mathcal{G}$, which satisfy
\begin{equation}
    \mathcal{J}^{\rm opt} = \left(\gamma n_{\rm occ}(\sqrt{|g|^2-\left(\mathcal{J}^{\rm opt}\right)^2}+|g|)^{2N}\right)^{1/(2N+1)}.
\end{equation}
For small $N$, the ratio $\mathcal{G}/|g|$ is close to $2$, but it approaches $1$ as $N$ increases. Meanwhile, $\Delta \omega_r = \mathcal{J}$ approaches $|g|$ as well.

\begin{figure}[t]
    \centering
    \includegraphics[width=0.45\textwidth]{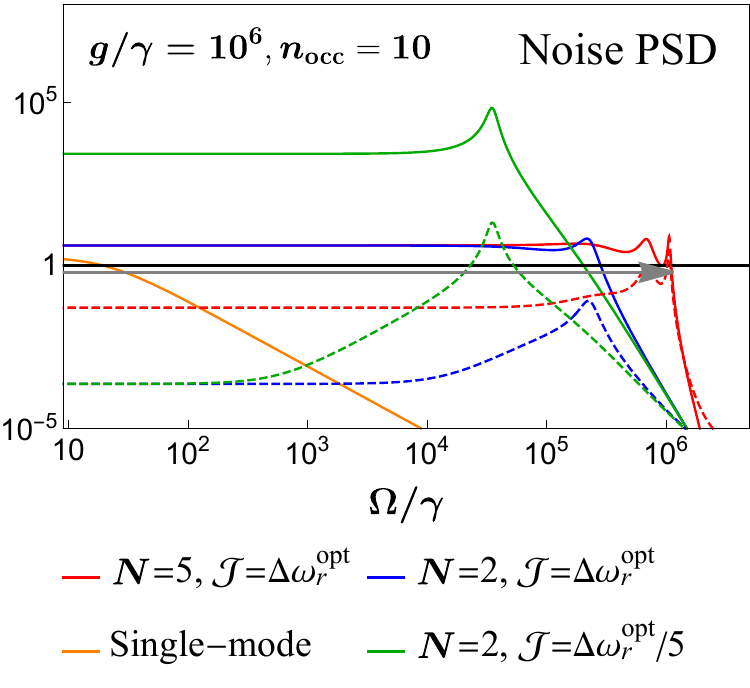}
    \caption{Noise PSDs for chain detectors with assumptions of $g/\gamma = 10^6$, $n_\occ = 10$, and $\gamma_r = \Delta \omega_{r}^{\rm opt}$. The black line represents readout noise, while the other solid lines depict the dominant contribution from $\hat{a}_0$, which surpasses the remaining intrinsic noise shown by the dashed lines. A decrease in $\mathcal{J}$ results in a squeezed spectrum. The gray arrow line highlights the range where Eq.~(\ref{eq:bw_nhc}) is applicable, specifically for $N=5$. For comparison, the intrinsic noise of a single-mode resonator is shown in orange.}
    \label{fig:PSDc}
\end{figure}

Figure~\ref{fig:PSDc} shows numerical examples of noise PSDs, with $\gamma_r = \Delta \omega_r^{\text{opt}}$ held constant and $\mathcal{J}$ varied. In these cases, the condition from Eq.~(\ref{eq:bw_nhc})—that the intrinsic noise in $\hat{a}_0$ (solid lines) surpasses other intrinsic noise sources (dashed lines)—is inherently satisfied. The figure illustrates that a smaller value of $\mathcal{J}$ has a minimal effect on $\Delta \omega_r$, even though it compresses the PSD within a narrower $\Omega$ range, thus validating our earlier argument. By setting $\gamma_r = \mathcal{J} = \Delta \omega_r^{\text{opt}}$, we achieve a relatively flat PSD within $\Delta \omega_r$, making it robust against potential variations in $\mathcal{J}$, especially considering the reasonable demands for the experimental dynamic range.

For varying $\mathcal{J}$, the PSD amplitude in the flat region can be derived as:
\begin{equation}
    |S_{0r}(\Omega=0)|^2n_{\rm occ} \approx 4n_\occ \gamma \gamma_r\mathcal{G}^{2N}/(\gamma_r\mathcal{J}^{N}+\mathcal{J}^{N+1})^2.
\end{equation}
which scales as $\propto \mathcal{J}^{-2N}$ when $\mathcal{J} \ll \gamma_r$. The PSD remains flat within $\Omega \ll \mathcal{J}$ and rapidly decreases outside of this range, as shown by the green line in Fig.~\ref{fig:PSDc}.

 \begin{figure}[t]
    \centering
    \includegraphics[width=0.45\textwidth]{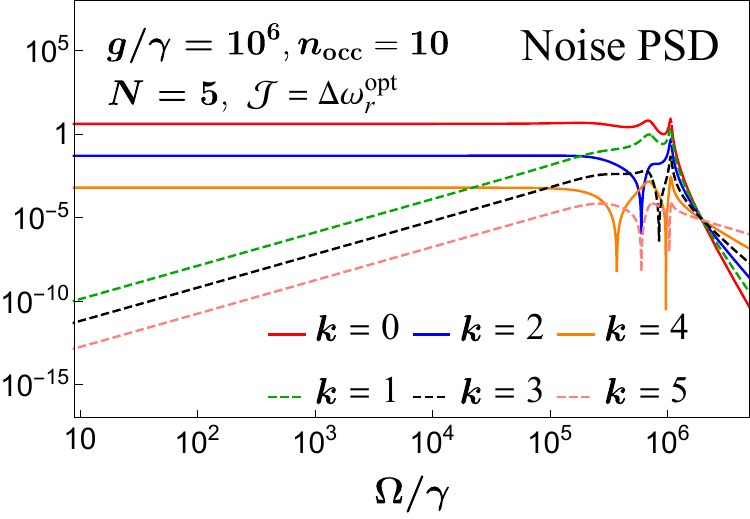}
    \caption{Contributions to the noise PSDs from different mode levels for the $N=5$ case shown in Fig.~\ref{fig:PSDc}. The solid lines represent the even-level contributions, which exhibit a nearly flat spectrum within the bandwidth, while the dashed lines correspond to the odd-level contributions, which scale as $\propto \Omega^2/\gamma_r^2$. The amplitudes of each class follow a geometric sequence with a ratio of $(\gamma_r/\mathcal{G})^4$, with the dominant contribution coming from the $N=0$ mode.}
    \label{fig:N=5PSDc}
\end{figure}

In Fig.~\ref{fig:N=5PSDc}, we show the contribution to PSDs from different levels for the $N=5$ case presented in Fig.~\ref{fig:PSDc}. Here, the relatively flat PSDs come from even-level modes, while the power-law dependent PSDs, scaling as $\propto \Omega^2/\gamma_r^2$, arise from odd-level modes. The amplitudes of each class follow a geometric sequence with a ratio of $(\gamma_r/\mathcal{G})^4$, with the dominant contribution from the $N=0$ mode. Within the response width, their PSDs can be approximated as:
\begin{equation}
    |S_{kr}|^2 n_\occ=\frac{\gamma_r^{2k}}{\mathcal{G}^{2k}}\times\left\{\begin{aligned}4, \qquad \quad &\text{for } N \in \text{ even}\\\,(k+1)^2\frac{\Omega^2}{\gamma_r^2}.\quad &\text{for } N \in  \text{ odd}
    \end{aligned}\right.
\end{equation}
The total contribution from odd-level modes is negligible compared to the even ones.

Higher-order contributions to noise can be accounted for by introducing an effective thermal occupation number, $n_{\rm occ}^{\rm eff}$, where the ratio $n_{\rm occ}^{\rm eff} / n_{\rm occ}$ represents the total noise PSD amplitude relative to the one from the $0$-th mode only. This is given by:
\begin{equation}
    |S_{0r}|^2n_{\rm occ}^{\rm eff} = \sum_{i=0}^{N}|S_{ir}|^2 n_\occ.
\end{equation}
Using the geometric series with a ratio of $(\gamma_r/\mathcal{G})^4$ from the even-level modes, we obtain:
\begin{equation}
    n_{\rm occ}^{\rm eff}  = n_\occ\frac{\mathcal{G}^4-(\gamma_r/\mathcal{G})^{4[N/2]}\gamma_r^4}{\mathcal{G}^4-\gamma_r^4}.
\end{equation}

These higher-order contributions can limit the optimal number of levels. According to Eq.~(\ref{snrS}), the sensitivity now becomes $\text{SNR}^2 (\overline{\omega}_\Psi)\propto \Delta \omega_r^{\text{opt}}/(n_{\rm occ}^{\textrm{eff}})^2$. Since both $\Delta \omega_r^{\rm opt}$ and $n_{\rm occ}^{\rm eff}$ increase with $N$, one can numerically solve for the optimal $N$ under a given $|g|/(\gamma n_{\rm occ})$. The optimal value of $N$ is always an odd number, which provides a larger response width $\Delta \omega_r^{\text{opt}}$ compared to the next smaller even number, with $n_{\rm occ}^{\rm eff}$ remaining nearly unchanged. A higher value of $g/(\gamma n_\occ)$ typically leads to a higher optimal number of levels, as the ratio of the geometric series becomes suppressed, i.e., $(\gamma_r/\mathcal{G}) \propto (\gamma n_{\rm occ}/g)^4$.

For the examples shown in Figs.~\ref{fig:PSDc} and \ref{fig:N=5PSDc} with $g/\gamma = 10^6$ and $n_{\rm occ} = 10$, the optimal level is $N = 9$, while for SRF detection with $g/\gamma = 10^{12}$ and $n_{\rm occ} = 100$, the optimal number is $N = 17$. At these optimal levels, the response width for both cases is greater than $80\%$ of the width when $N \rightarrow \infty$.

The binary tree scenario, as introduced in Sec.~\ref{sec:binarytree}, along with numerical computations based on optimized conditions, exhibits the same scaling of the response width as demonstrated in Eq.~(\ref{eq:rwo})~\cite{Chen:2021bgy}.

Historically, the goal of achieving a broad bandwidth with a large response originated in the context of gravitational wave detection, particularly in the concept of the white-light cavity~\cite{1997OptCo.134..431W}, where an anomalous medium with negative dispersion was proposed to counteract phase shifts in the off-resonant region. In our case, the combined use of beam-splitter and non-degenerate parametric couplings functions analogously: the response is first significantly enhanced by the non-degenerate parametric coupling $|G|$, while the addition of beam-splitter interaction with comparable strength $|g| \sim |G|$ flattens the response by reducing the peak, as seen in the denominators of Eqs.~(\ref{eq:YN}, \ref{eomcij}). This mechanism leads to a sequential broadening of the response width as the signal propagates. Consequently, by extending the response width from an order of $n_{\text{occ}} \gamma$ to $2|g|$, multi-mode resonators dramatically enhance the scan rate beyond that of single-mode resonators, as illustrated in Fig.~\ref{fig:PSDc}.

\section{Realization of Multi-Mode Resonators}\label{sec:multi-mode realization}

This section demonstrates the realization of multi-mode resonators that feature significantly broadened response widths, focusing particularly on the beam-splitter-type and non-degenerate parametric interactions between modes, as derived in Sec.~\ref{sec:multi-mode} and illustrated in Fig.~\ref{fig:ch}. We will also discuss potential challenges associated with their implementation. Notably, within a multi-mode resonator setup, only one mode at the lowest level needs to function as the probing sensor, as discussed in Secs.~\ref{sec:detectors} and \ref{sec:signals}. The remaining auxiliary modes may be circuit modes that do not require pick-up loops and are not embedded in a magnetic field background, or they can be other resonant cavity modes within the same cavity.

One method to achieve the necessary couplings is through three-wave mixing, where one pumping mode's frequency matches the sum or difference of the other two, thereby enabling non-degenerate parametric and beam-splitter-type interactions, respectively. Feasible implementations of three-wave mixing include the use of a Josephson Ring Modulator (JRM)~\cite{Bergeal:2010}, Superconducting Nonlinear Asymmetric Inductive eLement (SNAIL)~\cite{Frattini_2017}, and DC-driven Josephson junction effects~\cite{PCLC}. Notably, the JRM structure has recently been utilized to demonstrate response width broadening in a $N=1$ resonator composed of two circuit modes~\cite{Jiang:2022vpm}. Thus, we use JRMs as the benchmark for constructing multi-mode resonators, as illustrated in the bottom panel of Fig.~\ref{fig:ch}.

\subsection{Multi-Mode Resonators Connected by Josephson Ring Modulators}\label{sec:JRMrealization}

The use of a JRM to connect two modes with a third pumping mode is illustrated in the bottom panel of Fig.~\ref{fig:ch}. Each JRM consists of four Josephson junctions, each with the same critical current $I_0$, labeled by $i = 1, 2, 3, 4$. This results in a total Hamiltonian given by~\cite{Bergeal:2010}:
\begin{equation}
    H_{\rm ch}=\sum_{k=1}^{N} H^k_{\rm JRM}=\sum_{k=1}^{N} \sum_{i=1}^4 -I_0 \varphi_0 \cos \delta_i^k,
\label{eq: Htot}
\end{equation}
where the sum over $k$ labels the JRM connecting the $(k-1)$-th and the $k$-th modes with Hamiltonian $H^k_{\rm JRM}$, and each junction contributes a cosine potential. $\varphi_0 \equiv 1/(2e)$ is the reduced flux quantum, where $e$ is the elementary charge. $\delta_i^k$ represents the phase difference across the $i$-th Josephson junction of the $k$-th JRM, with the direction of phase difference defined as counter-clockwise within the JRM, as indicated by the purple arrow in the figure.

The resonant modes $\Phi_{k-1}$ and $\Phi_k$ are related to $\delta_i^k$ through a basis transformation, as mandated by Kirchhoff's voltage law, resulting in:
\begin{equation}
\begin{aligned}
    \frac{\Phi_{k-1}}{\varphi_0} &= \frac{\delta_1^k+\delta_2^k-\delta_3^k-\delta_4^k}{2},
    \frac{\Phi_{k}}{\varphi_0} = \frac{-\delta_1^k+\delta_2^k+\delta_3^k-\delta_4^k}{2}, \\
    \frac{\Phi_{p}}{\varphi_0} &= \frac{\delta_1^k-\delta_2^k+\delta_3^k-\delta_4^k}{2}, 
    \frac{\Phi_{L}}{\varphi_0} =- \left(\delta_1^k+\delta_2^k+\delta_3^k+\delta_4^k\right). \\
\end{aligned}
\label{eq: phi_transform}
\end{equation}
Here, $\Phi_L$ represents the loop flux threading through the JRM, and $\Phi_p$ serves as the pumping mode for the three-wave mixing. In the bottom panel of Fig.~\ref{fig:ch}, solid and dashed arrows in red represent $\Phi_{k-1}$ and $\Phi_k$, respectively, while dashed arrows in cyan denote $\Phi_p$.

Substituting the basis transformation from Eq.~(\ref{eq: phi_transform}) into the Hamiltonian Eq.~(\ref{eq: Htot}) yields:
\begin{equation}
\begin{aligned}
    H^k_{\rm JRM}=&-4 I_0\varphi_0 \Big( \cos \frac{\Phi_k}{2\varphi_0} \cos \frac{\Phi_{k-1}}{2\varphi_0} \cos \frac{\Phi_{p}}{2\varphi_0} \cos \frac{\Phi_L}{4\varphi_0}  \\
    &+ \sin \frac{\Phi_k}{2\varphi_0} \sin \frac{\Phi_{k-1}}{2\varphi_0} \sin \frac{\Phi_{p}}{2\varphi_0} \sin \frac{\Phi_L}{4\varphi_0} \Big)\\
    \approx&\frac{\sqrt{2}I_0}{4\varphi_0} \left[\left(\Phi_k^2+\Phi_{k-1}^2+\Phi_p^2\right) -\frac{\Phi_k\Phi_{k-1}\Phi_p}{\varphi_0}\right].
    \label{eq: HJRM}
\end{aligned}
\end{equation}
We universally set the loop flux threading through the JRM as $\Phi_L=\pi \varphi_0$. In the last line, we consider the leading order expansions where $\Phi_k, \Phi_{k-1}, \Phi_{p} \ll \varphi_0$. This approximation requires operation in the weak signal and low noise limit, where the flux in the resonant modes, $\Phi_k$ and $\Phi_{k-1}$, is significantly less than $\varphi_0$. To activate $\Phi_p$ as a pump mode, a current source labeled  $I_{g,G}$, illustrated in Fig.~\ref{fig:ch}, is injected at the bottom of the JRM, resulting in:
\begin{equation}
\begin{aligned}
    \Phi_{p}=  \frac{\sqrt{2}\varphi_0}{I_0} I_{g, G}, \qquad
    I_{g, G} \equiv I_g  \psi_{-}^k + I_G  \psi_{+}^k,
    \label{eq: pump}
\end{aligned}
\end{equation}
where $\psi_{\pm}^k \equiv \cos[(\omega_{\rm rf}^{k-1} \pm \omega_{\rm rf}^k)t]$, and the current amplitudes $I_{g}$ and $I_G$ are set to values smaller than $I_0$. These terms carry the frequencies of the sum and difference between the two resonant modes connected, simultaneously generating beam-splitter-type and non-degenerate parametric interactions.

Taking the currents from Eq.~(\ref{eq: pump}) into Eq.~(\ref{eq: HJRM}), the three-wave mixing term proportional to $\Phi_k \Phi_{k-1} \Phi_p$ facilitates both the beam-splitter-type and non-degenerate parametric interactions as described in Eq.~(\ref{eq:lag_ch}). Under the rotating wave approximation (RWA), the coupling strengths are given by:
\begin{equation}
\begin{aligned}
    |g|=  \frac{I_0}{4\varphi_0} \tilde{I}_g \kappa_k \kappa_{k-1} , \quad
    |G|=  \frac{I_0}{4 \varphi_0} \tilde{I}_G \kappa_{k} \kappa_{k-1},
\end{aligned}
\label{eq: gG_strength}
\end{equation}
where $\tilde{I}_g \equiv I_g/I_0$, $\tilde{I}_G \equiv I_G/I_0$, and $\kappa_k$, $\kappa_{k-1}$ are the zero-point uncertainties of the respective resonators, defined in Eq.~(\ref{eq: kappa}).

Note that in addition to the desired couplings, the first term in the last line of Eq.~(\ref{eq: HJRM}) results in frequency shifts $\delta \omega_{\rm rf}^k$ on the two resonant modes, expressed as
\begin{equation}
    \delta \omega_{\rm rf}^k =\frac{\sqrt{2} I_0 }{4\varphi_0} \kappa_k^2.
\end{equation}
To avoid disrupting the resonator's function, these shifts should remain smaller than the resonant frequencies $\omega_{\rm rf}^k$.

In practice, the amplitudes $|g|$ and $|G|$ dictate the response width $\Delta \omega_r$, as explored in Sec.~\ref{sec:multi-mode}. Their magnitudes can be approximated as:
\begin{equation}
    |g|,|G| \sim  I_0 \tilde{I}_{g/G} \kappa_k \kappa_{k-1}/\varphi_0 \sim  e^2 \tilde{I}_{g/G} \kappa_k \kappa_{k-1} E_J,
    \label{eq:gGEj}
\end{equation}
where $E_J \equiv I_0 \varphi_0$ denotes the Josephson energy. Achieving $E_J$ significantly higher than $\mathcal{O}(1)$\,GHz has been demonstrated in experiments~\cite{PhysRevB.63.064502,cite-key}, ensuring that both $|g|$ and $|G|$ can be engineered to be comparable to, but not exceeding, $\omega_{\rm rf}^k$ due to the consideration of frequency shifts. We will adopt $\Delta \omega_r \approx 2|g| \approx \omega_{\rm rf}$ of the probing sensor as the operational limit for multi-mode resonators. Before discussing the physics reach, we briefly outline several challenges to realization and experimental operation.

\subsection{Potential Challenges}

The implementation of three-wave mixing, as discussed above, has been previously demonstrated in Ref.~\cite{Bergeal:2010}. Additionally, a prototype that realizes both $g$ and $G$ types of couplings simultaneously using a Josephson Ring Modulator (JRM) has shown the potential to amplify a mimicking axion signal, achieving a 5.6-fold enhancement in SNR$^2$ with $N=1$~\cite{Jiang:2022vpm}. However, constructing a multi-mode resonator capable of significantly broadening the response bandwidth presents several challenges, which we will briefly address along with potential solutions below.

\begin{itemize}

    \item The foremost challenge is the compatibility of Josephson junctions with strong magnetic fields. Josephson junctions, being superconducting elements, require isolation from background magnetic fields to maintain superconductivity. One approach involves using a transmission line to connect the cavity or circuit with the Josephson junctions, thereby shielding them from direct magnetic exposure~\cite{Wurtz:2021cnm}. Moreover, the operation of Josephson junctions in magnetic fields up to $\mathcal{O}(1)\,\mathrm{T}$ has been successfully demonstrated~\cite{Rokhinson:2012dp}. Thus, the apparent incompatibility between Josephson junctions and strong magnetic fields can be effectively mitigated.

     \item Similarly, the operating temperature of the JRM must be kept below a certain threshold, typically ensuring that the thermal occupation number is dominated by quantum fluctuations. In the benchmark examples shown in Figs.~\ref{fig:PSDc} and \ref{fig:N=5PSDc}, we assume $n_{\rm occ} = 10$ for illustration. However, the environmental temperature surrounding the JRMs may differ from that inside the resonators. For instance, SRF cavities are typically cooled to around $1.8\,\mathrm{K}$ using liquid helium immersion~\cite{SHANHE:2023kxz}. Given that the setup is considerably larger than the cavity itself, there is enough space to accommodate an additional small dilution refrigerator, which can provide the mK-level temperatures required for JRM operation. Alternatively, the SRF cavity could be placed in the K-level stage of a sufficiently large dilution refrigerator to provide adequate cooling power, while the JRM is positioned in the mK stage of the same refrigerator to ensure proper operation.

         \item Additionally, when connections are made via JRMs and transmission lines, there is a potential for increased dissipation, which could lower the quality factor of the probing sensors. Consequently, precise noise control and calibration of $\gamma$ are necessary.

    \item Another potential challenge is the phase fluctuation of the pump mode $\Phi_p$, which can deviate from the frequencies at $\omega_{\rm rf}^{k-1} \pm \omega_{\rm rf}^k$. In the non-Hermitian chain model discussed in Sec.~\ref{sec:nhc}, the resulting phase fluctuations, denoted $\varphi^{g/G}_k$, can quench the amplification~\cite{PhysRevB.96.104503}. Misalignment of these phases can suppress the amplification of the quadratures, as described in Eq.~(\ref{eq: redef_XY}).
    
    For the non-Hermitian chain structure, phase misalignment can be recalibrated~\cite{Jiang:2022vpm}. The phase of the JRM can be adjusted by manipulating the phases of the drive currents that generate the beam-splitter-type and non-degenerate parametric interactions (i.e., the pump current $I_{g/G}$). Since phase alignment maximizes the output power, each JRM’s phase can be tuned during calibration to optimize this. This power maximization criterion can make the calibration process automatic.
     
     Typically, the JRM phase remains stable for much longer than the duration of the typical data acquisition cycle. The total integration time can be divided into several cycles, with calibration of all relevant parameters performed before data collection in each cycle.

    On the other hand, the binary tree model in Sec.~\ref{sec:binarytree} completely circumvents this issue, as both quadratures are equally amplified.

    \item In implementing three-wave mixing as outlined in Eq.~(\ref{eq: Htot}), with pumping modes defined in Eq.~(\ref{eq: pump}), higher-order terms such as $\Phi_k^2 \Phi_p^2$ and $\Phi_{k-1}^2 \Phi_p^2$ could potentially destabilize the resonator. The pumping mode $\Phi_p^2$ oscillates at frequencies $2\omega_{\rm rf}^k$ or $2\omega_{\rm rf}^{k-1}$, resulting in terms under the RWA that involve $\hat{a}_k^2$ and $\hat{a}_{k-1}^2$~\cite{Ruddy:2023wxc}:
    \begin{equation}
    \qquad H^k_{\rm JRM} \supset \frac{\sqrt{2}I_0}{32 \varphi_0} \tilde{I}_g \tilde{I}_G \left( \kappa_k^2 \hat{a}_k^2 + \kappa_{k-1}^2 \hat{a}_{k-1}^2 +\mathrm{c.c.}\right),
    \label{eq: SMS}
    \end{equation}
    which are known as the self-squeezing terms. These effects can be suppressed by detuning the pumping frequencies from $\omega_{\rm rf}^{k-1} \mp \omega_{\rm rf}^k$~\cite{Ruddy:2023wxc}. However, a slight deviation from the optimal setting could destabilize the system, particularly when $|g|$ and $|G|$ are large. Notably, in alternative configurations like the binary tree structure discussed in Sec.~\ref{sec:binarytree}, self-squeezing issues are naturally avoided by using only one type of coupling in each connection, facilitated by the introduction of auxiliary modes.

    \item Another source of nonlinearities arises from the expansion in Eq.~(\ref{eq: HJRM}), which requires that $\Phi_k \ll \varphi_0$ to prevent nonlinear effects from JRMs from invalidating our derivation. As shown in Eq.~(\ref{def:Phi}), $\Phi_k$ can be approximated by $\Phi_k \sim \langle \hat{a}_k^{\dagger} \hat{a}_k\rangle^{1/2} \kappa_k$. The steady-state energy in the resonators, from $\hat{a}_0$ to $\hat{a}_N$, grows geometrically with the ratio $(\mathcal{G}/\gamma_r)^2 > 1$. Therefore, we only need to ensure that the energy in the last mode, $\hat{a}_N$, satisfies $\Phi_N \ll \varphi_0$.

    The steady-state energy of $\hat{a}_N$ can be estimated as:
    \begin{equation}
        \langle \hat{a}_N^{\dagger} \hat{a}_N\rangle 
        \approx \frac{1}{\gamma_r } \Delta \omega_r \, |S_{0r}(\Omega=0)|^2  n_{\rm occ} \approx 4 ,
        \label{eq: energy_N}
    \end{equation}
    where we assume that $|S_{0r}|^2$ is roughly flat within the response bandwidth to simplify the expression, and we use the benchmark response width from Eq.~(\ref{eq:rwo}) in the final estimate.

    Meanwhile, from Eqs.~(\ref{eq: kappa}) and (\ref{eq: kappa_LC}), we see that zero-point fluctuations $\kappa_k$ can be adjusted by modifying the readout antenna structure in a cavity or by manipulating the inductance and capacitance of an LC circuit. This allows $\kappa_k$ to be reduced, ensuring that $\Phi_k \ll \varphi_0 \approx 1.65$ in natural units. However, excessively reducing $\kappa_k$ would decrease the magnitudes of $|g|$ and $|G|$ (as defined in Eq.~(\ref{eq:gGEj})), necessitating a higher Josephson energy $E_J$ to compensate.

    \item JRMs may experience power saturation when the injected power is too large. Since the power injected from the resonator $\hat{a}_k$ into the JRM is approximately given by $g\, \omega_{\rm rf}^k \langle \hat{a}_k^{\dagger} \hat{a}_k \rangle $, and the coupling strength $g$ between each resonator remains nearly constant, it suffices to ensure that the JRM connected to $\hat{a}_N$ does not reach saturation.

    In the benchmark case, with $\langle \hat{a}_N^{\dagger} \hat{a}_N\rangle \approx 4$ (from Eq.~(\ref{eq: energy_N})), and considering $\omega_{\rm rf}^N \approx 2\pi \,\mathrm{GHz}$ and $g \approx \omega_{\rm rf}^N$, the maximum injected power is approximately    $g\,\omega_{\rm rf}^N \langle \hat{a}_N^{\dagger} \hat{a}_N \rangle \approx -107.8\,\mathrm{dBm}$. 
    According to Ref.~\cite{Bergeal:2010vtg}, the saturation power of the JRM can reach around $-105\, \mathrm{dBm}$. Therefore, in the benchmark case considered in this study, the JRMs will not experience power saturation.

    \item The optimized response width discussed in Sec.~\ref{sec:multi-mode} assumes that the intrinsic noise levels across all modes are comparable, allowing the contribution from the lowest mode to dominate. This assumption may not hold if the auxiliary modes have lower quality factors $Q_{\rm int}$ than the probing sensor. To maintain dominance of the lowest mode, it is necessary that $(|g| + |G|)^2 \gamma_0 n_{\text{occ}}^0 > \Delta \omega_r^2 \gamma_1 n_{\text{occ}}^1$, where $\gamma_{0/1}$ and $n_{\text{occ}}^{0/1}$ represent the dissipation coefficients and thermal occupation numbers of the probing sensor and the auxiliary mode, respectively. This condition becomes particularly challenging in an SRF cavity with a high $Q_{\rm int}$ and low $\gamma_0$, necessitating that the auxiliary modes also maintain low $\gamma_1 n_{\text{occ}}^{1}$. To achieve this, one might consider employing other SRF cavities or different cavity modes within the same cavity as auxiliary modes.

  \item  The readout system is connected only to the final, highest-level mode in the network. This connection—typically implemented via an antenna or transmission line—is essentially identical to the standard readout approach used in single-mode resonators. However, a key distinction in the multi-mode setup is the significantly broadened response width, which requires the amplification chain to perform reliably across a wide frequency range. This imposes practical constraints on the choice of amplifiers. In particular, traveling-wave parametric amplifiers are well-suited for this application, as they offer broadband operation while maintaining quantum-limited noise performance.

\item In the binary tree configuration, achieving the full coherent enhancement of the signal—potentially improving the SNR$^2$ by a factor of $2^N$—requires phase matching across all probing sensors at the lowest level. Several factors can affect the relative phases of signals arriving at the readout port. First, the intrinsic phase coherence of the background source plays a key role. For ultralight DM, the coherence length is typically $\sim 10^3$ times longer than the Compton wavelength, which is itself comparable to or larger than the detector scale. This ensures phase coherence across the sensor array. A similar argument applies to HFGW detection using SRF cavities or LC circuits, where the GW wavelength is much longer than the detector dimensions. In the case of heterodyne upconversion detection, the background electromagnetic field used for mixing introduces an additional phase contribution. Furthermore, spatial separations between sensors can introduce relative phase shifts.

A viable calibration strategy is to inject a known coherent signal into all probing sensors simultaneously and adjust the relative phases by tuning the lengths of superconducting delay lines, thereby maximizing the resulting signal at the readout port.
\end{itemize}

\section{Simultaneous Resonant and Broadband Detection}\label{sec:resonant-broadband}
As demonstrated in Sec.~\ref{sec:multi-mode} and Sec.~\ref{sec:multi-mode realization}, the response width $\Delta \omega_r$ covered by each scan can be of the same order as $\omega_{\rm rf}$ for a multi-mode resonator. In this section, we apply this broadened width to the various detection schemes mentioned previously in Sec.~\ref{sec:signals}, establishing how the physics reach can be significantly deepened through multi-mode generalizations.

For both axion and HFGW detection using static magnetic fields, as well as dark photon detection without background fields, the reachable source frequency $\overline{\omega}_\Psi$ typically lies within the bandwidth $\Delta \omega_r$ centered around $\omega_{\rm rf}$. For these cases, Eq.~(\ref{snrS}) serves as a reliable approximation for estimating the sensitivity reach, with the integration time $t_{\text{int}}$ for each scan extending up to $t_e$ as $\Delta \omega_r$ approaches $\omega_{\text{rf}}\sim\overline{\omega}_\Psi$. Compared to single-mode resonators, multi-mode systems exhibit enhanced sensitivity. This enhancement is quantified by the ratio of their respective $\Delta \omega_r$, where for single-mode resonators it is $2 n_{\text{occ}} \gamma$ for LC circuits and SRF cavities, and $3\gamma$ for traditional cavities, resulting in:
\be
\frac{\textrm{SNR}^2_{\rm MM}}{\textrm{SNR}^2_{\rm SM}} = \frac{\Delta \omega_r^{\text{MM}}}{\Delta \omega_r^{\text{SM}}} \sim
\frac{Q_{\rm int}}{n_{\rm occ}}.\label{SNRRCC}
\ee
Here, `MM' and `SM' denote multi-mode and single-mode, respectively.

From Eq.~(\ref{SNRRCC}), we can generalize the SNR$^2$ calculations from single-mode resonators in Eqs.~(\ref{eq:SNRaSM}, \ref{eq:SNRDPSM}, \ref{SNRL}) to multi-mode limits, expressed as follows:\\
\\
$\bullet$ Axion:
      \be\begin{split}
{\rm LC\ Circuit}& : 
   \frac{\pi}{2} g_{a\gamma}^4 m_a \rho_{\text{DM}}^2 \eta^4 B_0^4 V^{{10}/{3}} Q_a Q_{\text{int}}^2 t_e/T^2.\\
  {\rm Cavity}& :
 2\pi \frac{g_{a\gamma}^4}{m_a^5} \rho_{\text{DM}}^2 \eta^4 B_0^4  V^2 Q_a Q_{\text{int}}^2 t_e,
 \label{eq:SNRaNN}
\end{split}\ee
 $\bullet$ Dark Photon:
\begin{align}
 {\rm LC\ Circuit}& :\frac{\pi}{2} \epsilon^4 m_{A^\prime}^{5} \rho_{\text{DM}}^2 \eta^4  V^{{10}/{3}} Q_{A^\prime} Q_{\text{int}}^2 t_e/T^2,\nn\\
{\rm Cavity}& : 2\pi  \frac{\epsilon^4}{m_{A^\prime}} \rho_{\text{DM}}^2 \eta^4  V^2 Q_{A^\prime} Q_{\text{int}}^2 t_e,\label{SNRDPMM}\\
{\rm SRF^{EM}}& :\frac{\pi}{2}  \epsilon^4m_{A^\prime} \rho_{\text{DM}}^2 \eta^4   V^2 Q_{A^\prime} Q_{\text{int}}^2 t_e/T^2.\nn
\end{align}
  $\bullet$ HFGW:
      \be\begin{split}
{\rm LC\ Circuit}& :\frac{1}{8\pi} h_0^4 \omega_h^9 \eta^4  B_0^4 V^{{14}/{3}}  Q_h Q_{\text{int}}^2 t_e/T^2,\\
 {\rm Cavity}& :\frac{1}{2\pi} h_0^4 \omega_h^3 \eta^4  B_0^4 V^{{10}/{3}}  Q_h Q_{\text{int}}^2 t_e.
 \label{eq:SNRHFGWMM}
\end{split}\ee
We utilize the same benchmark experimental parameters as those used for single-mode resonators, detailed in Sec.~\ref{sec:single-mode}. In Fig.~\ref{fig:bound}, we illustrate the multi-mode physics reach using dashed lines, requiring that Eqs.~(\ref{eq:SNRaNN}, \ref{SNRDPMM}, \ref{eq:SNRHFGWMM}) reach an SNR of $1$. Notably, Eq.~(\ref{SNRRCC}) demonstrates a significant enhancement in ${\rm SRF^{EM}}$ detection of dark photons, attributable to a high quality factor, $Q_{\rm int} = 10^{12}$.

\begin{figure}[htb]
    \centering
   \includegraphics[width=0.48\textwidth]{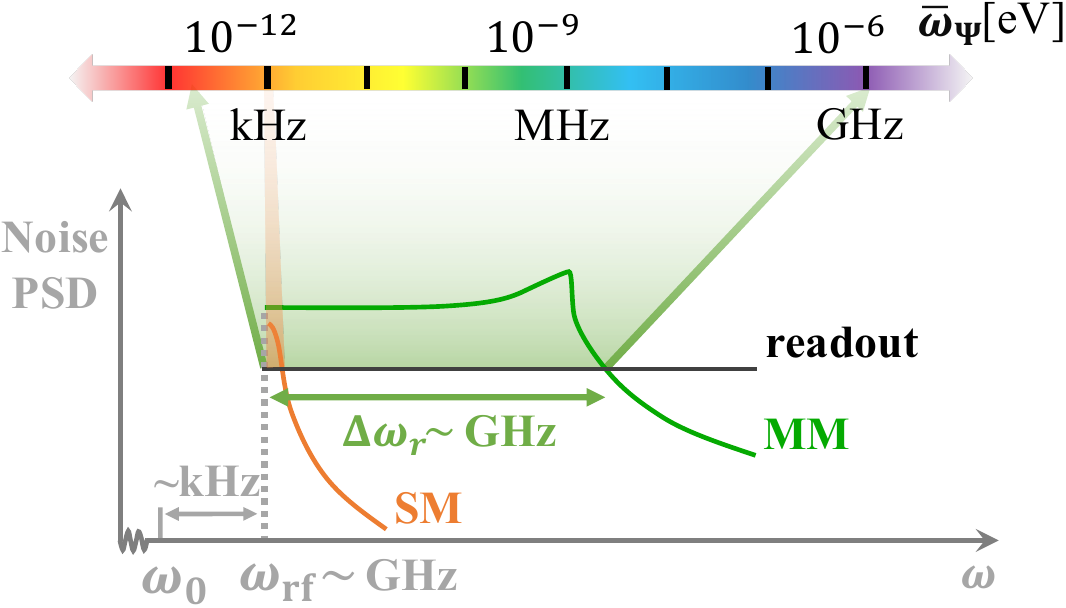}
   \caption{A schematic plot illustrating the response width, $\Delta \omega_r$, for single-mode (SM) and multi-mode (MM) generalizations of heterodyne upconversion detection. The corresponding coverages of the source frequency, $\overline{\omega}_\Psi$, are shown with shaded areas in orange for SM and green for MM, respectively. The noise PSDs follow the definitions outlined in Fig.~\ref{fig:PSDc}.}
    \label{fig:BW}
\end{figure}

On the other hand, for heterodyne upconversion-type detection, operating a pump mode with frequency $\omega_0$ in an SRF cavity allows for the excitation into a signal mode at $\omega_{\rm rf} = \omega_0 + \overline{\omega}_\Psi$, where $\overline{\omega}_\Psi$ can be significantly lower than $\omega_{\rm rf}$. Notably, when employing the multi-mode extension to SRF cavities, a wide range of $\overline{\omega}_\Psi$ spanning several orders of magnitude can be covered in a single scan. For example, setting $\omega_{\rm rf} - \omega_0 = 2\pi$\,kHz allows probing up to six orders of $\overline{\omega}_\Psi$, ranging from $2\pi$\,kHz to $\omega_{\rm rf} - \omega_0 + \Delta \omega_r \approx 2\pi$\,GHz, as illustrated in Fig.~\ref{fig:BW}. In principle, even lower frequencies could be explored by further reducing $\omega_{\rm rf} - \omega_0$, although this would introduce more intrinsic noise~\cite{Berlin:2019ahk,Berlin:2020vrk}. The SNR can be estimated by setting $t_{\rm int} = N_e\, t_e$ in Eq.~(\ref{eq:SNR}), where $N_e = 6\ln10$ represents the number of $e$-folds between kHz and GHz. This yields the ratio:
\begin{equation}
\frac{\textrm{SNR}^2_{\rm HUMM}}{\textrm{SNR}^2_{\rm HUSM}} \simeq N_e\frac{\overline{\omega}_{\Psi}\,Q_{\rm int}}{\omega_{\rm rf}\,n_{\rm occ}},\label{SNRRSRF}
\end{equation}
where `HU' denotes heterodyne upconversion detection. This enhancement is particularly evident in SRF detection for axion DM and HFGW, thanks to the high-quality factor.

From Eq.~(\ref{SNRRSRF}), the SNR$^2$ expressions for the multi-mode generalization of heterodyne upconversion detection are as follows:
\\
\\
$\bullet$ Axion:
      \be\begin{split}
 {\rm SRF^{EM}}& : \frac{3\pi\,\text{ln10}}{4} \frac{g_{a\gamma}^4}{m_a} \rho_{\text{DM}}^2  \eta^4 B_0^4  V^2 Q_a Q_{\text{int}}^2t_e/(\omega_{\rm rf}^2T^2).\\\label{eq:SRFEMA}
\end{split}\ee
 
  $\bullet$ HFGW:
\begin{align}
 {\rm SRF^{EM}}& :   \frac{3\,\text{ln10}}{16\pi} h_0^4 \omega_h^7 \eta^4  B_0^4 V^{{14}/{3}}  Q_h Q_{\text{int}}^2 t_e\omega_{\rm rf}^2/T^2.\nn\\
 {\rm SRF^{mech}}& :  \frac{3\,\text{ln10}}{16\pi} \frac{h_0^4 \omega_h^7 \vert \eta_p^t\,\eta_p^h\,L_p(\omega_h)\vert^4 B_0^4 V^{2} Q_h Q_{\text{int}}^2 t_e \omega_{\text{rf}}^2}{T^2 N_M^2}. \nn\\ \label{SNRMM}
\end{align}
We depict the corresponding physics reach with dashed lines in Fig.~\ref{fig:bound}. The electromagnetic coupling cases of ${\rm SRF^{EM}}$ demonstrate significant enhancement across various frequencies. However, in the mechanical coupling case, ${\rm SRF^{mech}}$, there is negligible enhancement below $\mathcal{O}(100)$\,kHz, as the response width in the single-mode case can already cover the order of $\omega_h$ at these frequencies.

Notably, apart from the sensitivity enhancement described in Eq.~(\ref{SNRRSRF}), the multi-mode upgrade eliminates the need to tune $\omega_{\rm rf} - \omega_0$ for each scan step, resulting in a broadband detector. Compared to traditional broadband setups in Refs.~\cite{Kahn:2016aff,Berlin:2020vrk}, the multi-mode design significantly enhances the response to the signal, combining the advantages of resonant detection. In the standard SRF broadband setup, $\omega_0 = \omega_{\rm rf}$, and an over-coupled readout coupling $\gamma_{B}$ is employed, enabling the probing of $\overline{\omega}_{\Psi}/(2\pi)$ above $10$\,kHz in the off-resonant region~\cite{Berlin:2020vrk} with $|S_{0r}|^2 \approx 4\gamma\gamma_B/ \overline{\omega}_{\Psi}^2$. The corresponding signal and noise PSDs are
$S_{\text{sig}} \simeq \gamma_B \alpha^2 S_\Psi/\overline{\omega}_{\Psi}^2$ and $ S_{\text{noise}} \simeq 1$, respectively, leading to the following SNR ratio:
\be
    \frac{\text{SNR}^2_\text{HUMM}}{\text{SNR}^2_{\rm BB}} \simeq \frac{
    \overline{\omega}_{\Psi}^4\,  Q_{\rm int}^2} {16\, \gamma_{B}^2\, \omega_{\rm rf}^{2}\,n_\occ^{2}},\label{eq:SNRB}
\ee
where `BB' denotes broadband detection. The significant enhancement factor in Eq.~(\ref{eq:SNRB}) is primarily due to the severely suppressed off-resonant response in the standard broadband SRF. For other types of broadband searches, such as those using LR circuits proposed in Ref.~\cite{Kahn:2016aff}, the sensor consistently responds to the PSD of effective currents induced from bosonic fields. Consequently, the resultant SNR$^2$ is substantially lower than that achieved with heterodyne upconversion, reduced by a factor of $Q_{\rm int}^2/n_{\occ}^2$.

Note that the above discussion does not account for the potential coherent enhancement of SNR$^2$ by a factor of $2^N$ in the binary tree configuration. In principle, one could form a detector network by using multiple probing sensors placed at the lowest level of the tree.

\section{Discussion and Conclusions}\label{sec:conclusion}
This work demonstrates the efficacy of multi-mode resonators in achieving the advantages of both resonant and broadband detection. These resonators exhibit significant responses to signals across a sensitive bandwidth, spanning one or several orders in the frequency domain of the sources. This capability is achieved by employing both beam-splitter and non-degenerate parametric interactions to connect adjacent modes. By tuning these interactions to comparable magnitudes, off-resonant phases are effectively canceled sequentially, thereby enhancing both the bandwidth and the overall response of the system. Consequently, both the peak value and bandwidth of the scattering matrix increase sequentially during propagation towards the readout port, while the readout noise, which sets the standard quantum limit of the sensitive response width, remains unaffected. By upgrading to multi-mode detectors, the scan rate can be increased by a factor of approximately $Q_{\rm int}/n_\text{occ}$ compared to single-mode detectors. Moreover, the need for frequency tuning and calibration in single-mode resonators is eliminated, saving valuable time and enabling the scanning of large, unexplored regions of axion and dark photon DM, along with HFGWs, within a reasonably short timeframe. Notably, this includes the exploration of the well-motivated QCD axion~\cite{Peccei:1977hh} DM mass window above kHz.

The practical implementation of this concept relies on utilizing Josephson junctions, achievable with mature superconducting technology. The stability of the sensitive response width to variations in the two coupling values ensures the robustness of the quantum network. In the chain model described by Eq.~(\ref{eq:lag_ch}), calibration of the relative phases of the two couplings is necessary, and potential decoherence may arise from self-squeezing and phase fluctuations of the pumping modes~\cite{PhysRevB.96.104503}. However, such issues are circumvented in the binary tree model described in Eq.~(\ref{HBT}), where the two quadratures are equally amplified~\cite{Chen:2021bgy}. Another crucial consideration is the intrinsic dissipation of the probing sensors, which requires precise control once the multi-mode array is formed.

Note that the multi-mode resonators discussed in this work are compatible with squeezing technology employed at the readout port~\cite{Zheng:2016qjv, Malnou:2018dxn, HAYSTAC:2020kwv, Lehnert:2021gbj}. Both approaches aim to increase the range in which intrinsic noises dominate over readout noise. To further enhance sensitivity, additional probing sensors can be incorporated~\cite{Chen:2021bgy, Brady:2022bus}, which can be naturally embedded into a multi-mode network like the binary tree. Utilizing spatially distributed sensors and sensors with different sensitive directions can reveal both macroscopic properties and the microscopic nature of potential sources, such as angular distribution and polarization~\cite{Foster:2020fln, Chen:2021bdr}.

\hspace{5mm}
\begin{acknowledgements}
We are grateful to Raffaele Tito D'Agnolo, Nick Houston, Minyuan Jiang, Yonatan Kahn, Xuegang Li, Zhen Liu, Yiqiu Ma, Jan Sch$\ddot{\text{u}}$tte-Engel, Tao Shi, Junhua Wang, and Bin Xu for useful discussions.
This work is supported by the National Key Research and Development Program of China under Grant No. 2020YFC2201501. 
The Center of Gravity is a Center of Excellence funded by the Danish National Research Foundation under grant No. 184.
Y. C. acknowledge support by VILLUM Foundation (grant no. VIL37766) and the DNRF Chair program (grant no. DNRF162) by the Danish National Research Foundation, the European Union’s H2020 ERC Advanced Grant “Black holes: gravitational engines of discovery” grant agreement no. Gravitas–101052587, the Rosenfeld foundation in the form of an Exchange Travel Grant and by the COST Action COSMIC WISPers CA21106, supported by COST (European Cooperation in Science and Technology).
Views and opinions expressed are however those of the author only and do not necessarily reflect those of the European Union or the European Research Council. Neither the European Union nor the granting authority can be held responsible for them. This project has received funding from the European Union's Horizon 2020 research and innovation programme under the Marie Sklodowska-Curie grant agreement No 101007855 and No 101131233. J.S. is supported by the National Key Research and Development Program of China under Grants No. 2020YFC2201501 and No. 2021YFC2203004, Peking University under startup Grant No. 7101302974, the NSFC under Grants No. 12025507, No. 12150015, No. 12450006.
\end{acknowledgements}

\end{document}